\documentstyle[epsfig,referee]{mn2e}
\newcommand{\ddd}{\,\mathrm{d}}

\title[A far infrared view of the Lockman Hole] 
{A far infrared view of the Lockman Hole from ISO 95 $\mu$m observations -  
I. A new data reduction method \thanks{Based on observations obtained with 
the {\sl Infrared Space Observatory}, an ESA science missions with instruments
and contributions funded by ESA Member States and the USA (NASA).}} 

\author[G. Rodighiero, C. Lari, A. Franceschini, A. Gregnanin and D. Fadda] 
  {G.~Rodighiero$^1$,\thanks{E-mail: rodighiero@pd.astro.it} 
  C.~Lari$^2$, A.~Franceschini$^1$, A.~Gregnanin$^1$,
  D.~Fadda$^3$$^{,4}$\\ 
  $^1$ Dipartimento di Astronomia, Universit\`a di Padova, 
 Vicolo dell'Osservatorio 2, I-35122 Padova, Italy \\ 
  $^2$ Istituto di Radioastronomia del CNR, via Gobetti 101, I-40129 Bologna, Italy \\ 
  $^3$ Caltech, SIRTF Science Center, MC 220-6, Pasadena, CA 91126, USA\\
  $^4$ Instituto de Astrof\'\i{}sica de Canarias (IAC),  Via Lactea S/N, E-38200 La Laguna, Spain} 

\date{Released 2002, Dec 1} 
 
\pagerange{\pageref{firstpage}--\pageref{lastpage}} \pubyear{2002} 
 
\def\LaTeX{L\kern-.36em\raise.3ex\hbox{a}\kern-.15em 
    T\kern-.1667em\lower.7ex\hbox{E}\kern-.125emX}

\begin{document} 
 
\label{firstpage} 
 
\maketitle 
 
\begin{abstract} 

We report results of a new analysis of a deep 95 $\mu$m imaging 
survey with the photo-polarimeter ISOPHOT on board the Infrared
Space Observatory, over a 40$^\prime\times 40^\prime$ area within the Lockman Hole.
To this end we exploit a newly developed parametric algorithm able
to identify and clean spurious signals induced by cosmic-rays impacts
and by transient effects and non-linearities in the detectors.
These results provide us with the currently deepest -- to our knowledge --
far-IR image of the extragalactic sky. Within the survey area
we detect thirty-six sources with S/N$>3$ (corresponding 
to a flux of 16 mJy), making up a complete
flux-limited sample for $S_{95 \mu m} \geq 100$ mJy. Reliable
sources are detected, with decreasing but well-controlled completeness,
down to $S_{95 \mu m} \simeq 20$ mJy.
The source extraction process and the completeness, the photometric and
astrometric accuracies of this catalogue have been tested by us with 
extensive simulations accounting for all the details of the procedure.
We estimate source counts down to a flux of $\sim 30$ mJy, at which
limit we evaluate that from 10\% to 20\% of the cosmic IR
background has been resolved into sources (contributing to the CIRB
intensity $\simeq 2.0\ 10^{-9}\ W/m^2/sr$).

The 95 $\mu$m galaxy counts reveal a steep slope at $S_{95 \mu m} \le 100$ 
mJy  ($\alpha\simeq 1.6$), in excess of that expected for 
a non-evolving source population. 
The shape of these counts agrees with those
determined by ISO at 15 and 175 $\mu$m, and starts setting
strong constraints on the evolution models for the far-IR galaxy populations.

\end{abstract} 
 
 
\section{Introduction} \label{intro} 

The star formation in local galaxies is univocally found to be associated
with dense dust-obscured clouds which are optically thick in the UV and
become transparent or even emissive at long infrared wavelengths.
For this reason, the far infrared domain is expected to be quite instrumental 
for studying not only the physics of star formation in our and closeby 
galaxies,
but also the early phases of galaxy evolution, when stellar formation 
was far enhanced compared to what happens in the local universe.

The IRAS satellite mission proved indeed the potential of IR  
observations to detect galaxies optically obscured by dust. 
Only a fraction of the 25.000 sources detected in the All Sky Survey were 
found to have bright optical counterparts  
(Soifer et al., 1987), and of these most are local late-type spirals. 

The IRAS survey was devoted to investigate the properties of the IR
emission by local galaxies, at redshift $<$ 0.2 (Ashby et al., 1996). 
Only few sources were detected by IRAS at higher redshifts, typically ULIRGs 
magnified by gravitational lenses, like F10214+4724 (z=2.28, 
Rowan-Robinson et al., 1991). IR galaxy counts based on the IRAS data
(Rowan-Robinson et al., 1994; Soifer et al., 1984) 
showed some marginally significant excess of faint sources with respect to 
no evolution models (Hacking et al., 1987; Franceschini et al. 1988; Lonsdale et al.,
1995;  Gregorich et al., 1995; Bertin et al, 1997), but did not provided 
enough statistics and
dynamic range in flux to discriminate between evolutionary scenarios.

The cosmological significance of far-IR studies was first emphasized 
by the COBE detection of an isotropic far-IR/submillimeter background,  
of extragalactic origin (CIRB), interpreted 
as the integrated emission by dust present in distant and primeval galaxies 
(Puget et al., 1996, Hauser et al., 1998), including an energy density
larger than that in the UV/optical background (Lagache et al., 1999). 

With the advent of the Infrared Space Observatory (ISO, Kessler et al., 1996) 
the improved resolution and sensitivities of its cameras
made possible deeper IR surveys, allowing us for the first time to detect  
faint IR galaxies at cosmological distances, both in the mid and in the far infrared. 

The deep 15 $\mu$m counts determined with ISOCAM (Cesarsky et al., 1996) have revealed a 
very significant departure
from the euclidean slope (Elbaz et al., 1999; Gruppioni et al., 2002), 
which has been interpreted as evidence for a strongly evolving population 
of starburst galaxies (Franceschini et al., 2001; Chary \& Elbaz, 2001; Xu et 
al., 2001;  Oliver et al., 2001). 
ISO surveys at longer (far-IR) wavelengths with the photo-polarimeter
ISOPHOT (Lemke at el., 1996) found some evidence of evolution in the 
175 $\mu$m channel (the FIRBACK survey: Puget et al. 1999, Dole et al. 2001;
ELAIS survey, Efstathiou et al., 2000; the Lockman Hole survey: Kawara et al.
,1998, Matsuhara et al., 2000; Juvela et al., 2000).
Unfortunately, at such long wavelengths the ISO observatory was quite limited by 
source confusion to moderately faint fluxes ($S_{175 \mu} \ge 135$ mJy).

In principle, the shorter wavelength 95 $\mu$m C100 channel of ISOPHOT could
allow a substantial improvement, by a factor 2, in spatial resolution
and a significantly lower confusion noise.
Furthermore, the filter samples in an optimal way the dust emission peak
in the Spectral Energy Distribution (SED) of star-forming galaxies 
around 60 to 100 $\mu$m. 
For luminous infrared galaxies, emitting more than 80\% of the flux 
in the far-IR, this far-IR peak is the best measure of the  
bolometric luminosity of such galaxies, and the best estimator of their  
star formation rate. 

The well-known problem with ISOPHOT C100 observations was the 
difficulty in the data reduction to account for all the instrumental 
effects, due to cosmic-ray impacts and transient effects in the 
detectors producing spurious detections which may contaminate 
the final source lists. 

We present in this paper a new method for the reduction of ISOPHOT C100 data, 
which we developed along similar lines as the code designed for ISOCAM 
data reduction by Lari et al. (2001, hereafter L01). 
We illustrate the value of this method with application to a deep ISOPHOT 
C100 survey in the Lockman Hole, a region particulary suited for the 
detection of faint infrared sources due to its low cirrus emission. 
The good quality of these data and a careful 
reduction allow us to reach faint detection limits ($\sim 20$ mJy).  
We focus in particular on the implications for the evolutionary models as  
derived from galaxy counts. Our results favour a scenario dominated by 
a strongly evolving population, quite in agreement with the model discussed 
by Franceschini et al. (2001). 
In a forthcoming paper we will discuss the optical identifications of these
Lockman 95 $\mu$m sources with radio and ISOCAM mid-IR counterparts  
(Rodighiero et al., in preparation) and we will explore the nature of our 
far-IR sources. 
 
The present paper is organized as follows. 
In Section 2 we discuss our reduction technique. 
In Section 3 we comment on the simulations we used to compute the
completeness of our sample, and the photometric corrections.
Section 4 is devoted to the flux calibration.
We then report in Section 5 on our application to the Lockman Hole 95 $\mu$m
data and our results on source counts. Our conclusions are reported in Section 6.
In the three Appendices we detail some technical aspects of our data
reduction method.

We assume throughout this paper $\Omega_M$=0.3,  $\Omega_{\Lambda}$=0.7
and $H_0$=65 $Km~ s^{-1} Mpc^{-1}$.

\section{A NEW TOOL FOR THE REDUCTION OF ISOPHOT-C DATA}  \label{phot} 

The reduction of ISO data requires a careful treatment 
of various external and instrumental effects affecting the detectors.  
The results recently obtained by L01 in developing 
a data reduction technique for ISOCAM data prompted us to attempt
a similar approach for the analysis of ISOPHOT data. 
 
PHOT C100 is a $3\times3$ array of Ge:Ga with 0.7$\times$0.7$\times$1mm elements. 
The effective size of the pixels on the sky
is 43.5$\times$43.5 arcsec, the distance between the pixel centers is 
46.0 arcsec. There are 6 filters available for C100, covering the
wavelength range from $\sim$60 to $\sim$100 $\mu$m. 
 
As for the case of ISOCAM (Long Wavelength) Si:Ga detectors, 
two main effects must be considered when
dealing with ISOPHOT-C data, produced by cosmic ray impacts 
({\it glitches}) and 
detector hysteresis (i.e. the slow response of the detector to 
flux variations). 
The method discussed by L01 was based on the assumption that 
the incoming flux of charged particles generates transient 
behaviours with two different time scales: 
a fast ({\bf breve}) and a slow ({\bf lunga}) one. 
The method basically consists in looking at the time history of each detector pixel and 
identifying the stabilization background level. Then 
it models the glitches, the background and the sources with all  
the transients over the whole pixel time history. 
 
\begin{figure*} 
  \begin{center} 
    \epsfig{file=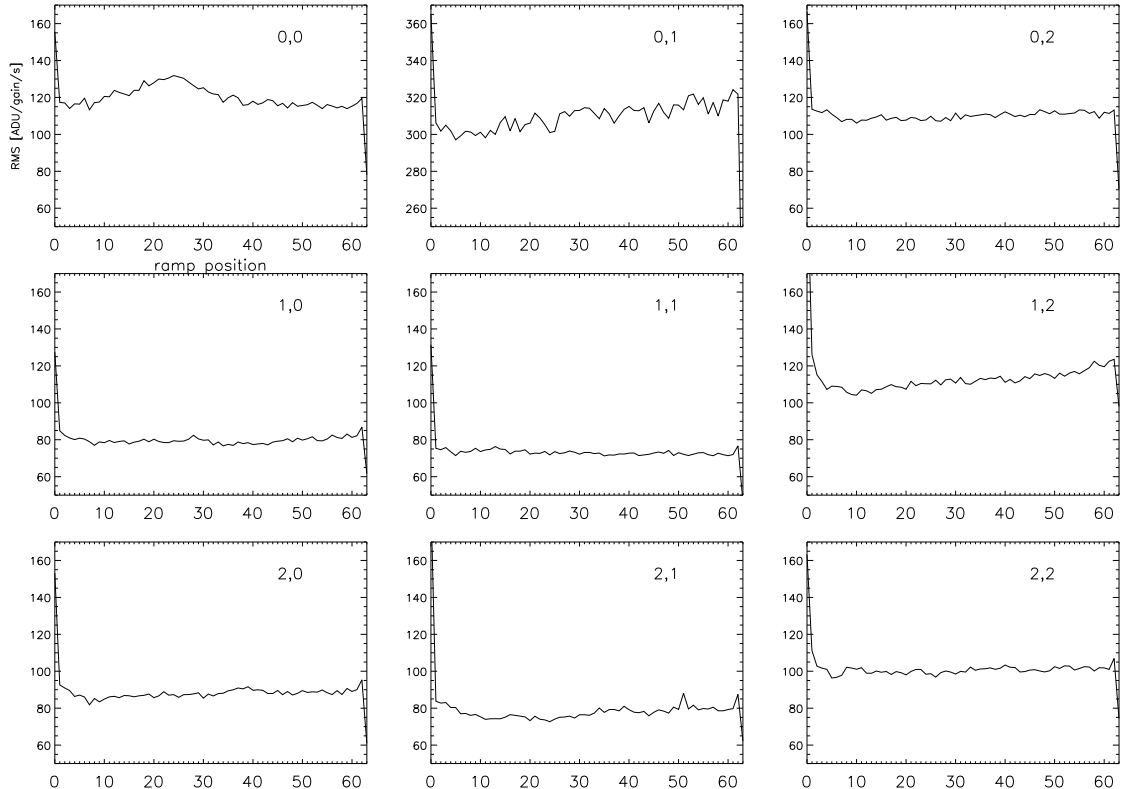, width=15cm} 
  \end{center} 
\caption{For each PHOT C100 detector pixel the RMS of the median 
signal over the whole pixel history in every ramp position is shown.  
The values are similar, except for pixel  
(0,1) which clearly shows a noise a factor 3 higher than the others. 
The data reported here come from an analysis of one ELAIS S1 raster  
observation, for which we have taken 
all the ramps along the time history and compute the median over each ramp 
position, excluding those readouts 
with a signal exceeding $|3\times \sigma|$.
Given that one ramp position is 1/32s, the 63 events reported on the x-axis
correspond to a temporal sequence of 2 seconds. } 
\label{dev_pix} 
\end{figure*}

If the approach is similar to that for ISOCAM, some peculiarities of the 
far-infrared detectors need to be treated with specific care. 
We found, in any case, that the description of transients with the equations 
used in the case of ISOCAM pixels provides good fits also to the
PHOT-C data (after adapting the temporal and charge parameters). 

Let us mention that a model like that of Fouks \& Schubert (1995) 
can fit, with suitable parameters, the brighest sources, both in the
case of CAM (Coulais \& Abergel, 2000) and of PHOT (Coulais et al., 2001),
indicating very similar shapes of the short term transient. 
In the case of CAM the method is an alternative to this model.
As far as PHOT is concrned, the glitches look very similar to those found
in the time histories of CAM's pixels.
Moreover, long term transients have been observed also in the PHOT 
detectors.
These considerations suggest that the Lari model is applicable also
to PHOT data, using the appropriate parameters.
We will show that, in spite of the numerous 
glitches recorded, it has been possible to define a reduction strategy 
with reliable results, including the stability of the temporal parameters.

As already mentioned, the method by Lari et al. (2001) 
describes the sequence of readouts, or time history,
of each pixel of CAM/PHOT detectors in terms of a mathematical model
for the charge release towards the contacts.
Such a model is based on the assumption of the existence, in each pixel,
of two charge reservoirs, a short-lived one $Q_b$ (breve) and a
long-lived one $Q_l$ (lunga), evolving independently with a different
time constant and fed by both the photon flux and the cosmic rays.
Such a model is fully conservative, and thus the observed signal $S$ is
related to the incident photon flux $I$ and to the accumulated charges
$Q_b$ and $Q_l$ by the (see also Lari et al., 2002):
\begin{equation}
S = I - \frac{\ddd Q_{tot}}{\ddd t} = I - \frac{\ddd Q_b}{\ddd t} - \frac{\ddd Q_l}{\ddd t}~,
\end{equation}
where the evolution of these two quantities is governed by the same
differential equation, albeit with a different efficiency $e_i$
and time constant $a_i$
\begin{equation}
\frac{\ddd Q_i}{\ddd t} = e_i\,I - a_i\,Q_{i}^2~~~~~\mathrm{where}~~~i=b,l~,
\end{equation}
so that
\begin{equation}
S = (1-e_b-e_l)\,I + a_b\,Q_b^2 + a_l\,Q_l^2~.
\end{equation}
The values of the parameters $e_i$ and $a_i$ are estimated from the data and
are constant for a given detector, apart from the scaling of $a_i$ for the
exposure time and the average signal level along the pixel time history,
which is governed by the
\begin{equation}
a_i = \frac{t}{t_0}\,\sqrt{\frac{S}{S_0}}~a_{i,0}~,
\end{equation}
where $a_{i,0}$ is the value of $a_i$ relative to a reference exposure time
$t_0$ and average signal level $S_0$.
The model for the charge release, however, is exactly the same for
CAM and PHOT detectors.

The values for the model parameters of CAM and PHOT respectively are 
reported in Table \ref{param}, together with the times characteristic of the
long and the short transients (dt/a\_i).

\begin{table} 
\caption{Model parameters for CAM and PHOT.} 
\label{param} 
\begin{tabular}{|c|c|c|c|c|c|c|c|} 
\hline 
\hline 
 ~ & dt(sec) &  e\_l& e\_b &  a\_l & a\_b &  dt/a\_l (sec) &  dt/a\_b(sec)\\
\hline 
\hline     
CAM  &   2.5   &  0.45   & 0.1   &   0.107609  &  0.00634620  &  23.2323  &  393.937\\
\hline
PHOT  &  1/32  &  0.36   & 0.1	 &  0.00529595 & 0.000197009  &  5.90074  &  158.622\\
\hline 
\hline 
\end{tabular} 
\end{table}

\subsection{DATA REDUCTION} 
\label{red} 

The Interactive Data Language (IDL) has been used to develop all the  
procedures needed for the reduction of PHOT-C data.

Before discussing in detail the reduction procedures, let us  
note the main difference between ISOCAM and PHOT. The ISOPHOT pixels can be  
considered as independent detectors, with entirely uncorrelated behaviours
and responses. This is not the case for ISOCAM, where all pixels  
have a common electronics. An example is presented in Figure \ref{dev_pix}, 
where we report a statistically rich data-set from the ISOPHOT ELAIS surveys
(Oliver et al., 2000) in the southern S1 field.
For an ELAIS raster observation with ISOPHOT, 
we have taken all the ramps along the time history and computed the 
median over each ramp position (one ramp is composed of 64 readouts), 
excluding those readouts 
with a signal exceeding $3\times \sigma$ (where $\sigma$ is the standard 
deviation of the signal over the whole pixel history). 
In the figure we report for each PHOT C100 detector pixel the RMS of the median 
signal over the whole pixel history in every ramp position.  
This gives a quantitative idea of the intrinsic 
noise of each detector pixel. The values are similar, except for pixel  
(0,1),
which clearly shows a noise $\sim$ 3 times higher than the others.  
A similar analysis is performed over all ELAIS rasters (both in northern and 
southern fields) and found very similar levels for each pixel. This  
consideration allowed us to consider the pixels' responsivities  
stable as a function of time and of the orbital position.        
The different peculiarities of the nine ISOPHOT C100 pixels 
imply that any kind of correction must be computed as a function 
of the pixel.

\subsection{From Raw Data through the fitting algorithm} \label{red1} 

The raw data (ERD level) are converted into a raster structure containing 
instrumental informations on the observation, and astrometric informations  
on every pointing. 
The ramps (in Volts) are corrected for the non-linear response of the
detector using a new technique (Appendix A) and converted in ADU/gain/s. 
The standard PHOT Interactive Analysis (PIA, Gabriel \& Acosta-Pulido, 1999) 
package process the data fitting the ramps (in units of Volts). 
 
In our procedure, the data are corrected for short-time cosmic rays. Readouts affected  
by such events are masked and their positions stored before copying the 
``deglitched'' data into a new structure (called 'liscio', as for ISOCAM).  
 
We evaluate the general background as the stabilization level along the whole 
time history of each pixel (for clarity, in the following we will refer to  
this quantity as the ``stabilization background''). 
 
We then apply a constant positive offset signal to the data in order 
to take into account the contribution of thermal dark current (which 
is not otherwise accounted for in the preliminary pipeline) when the 
latter is estimated to be important, i.e.\ when the deepest dipper's 
depth exceeds 10\% of the stabilization background. 
 
The task computing the stabilization background also performs an initial 
guess of the fitting parameters, storing them in the 'liscio' structure. 
 
The signal as a function of time is finally processed, independently for  
every pixel. The fitting procedure models the transients along the time 
history, and the features on both short and long timescales produced by 
cosmic ray impacts.  
At this level the code estimates several quantities needed to build 
the final maps on which source extraction will be performed: 
 
\begin{itemize} 
\item 
the charges stored into the $breve$ and $lunga$ reservoirs at each readout. 
\item 
The local background, i.e. the signal to be expected on the basis of the 
previously accumulated charges if only the stabilization background were 
hitting the detector. 
\item 
The model signal produced by the incident flux coming from both the 
stabilization background and detected sources. 
\item 
The ``reconstructed'' signal, i.e. the model signal 
recovered not only from glitches, but also from the transients due 
to the pointing-to-pointing incident flux change.
\end{itemize} 
 
Furthermore the code recognizes sources (above a  
given threshold level) and recovers all the time histories ``reconstructing''  
the local background as it would appear in absence of glitches. 
 
\begin{figure*} 
  \begin{center} 
\epsfig{file=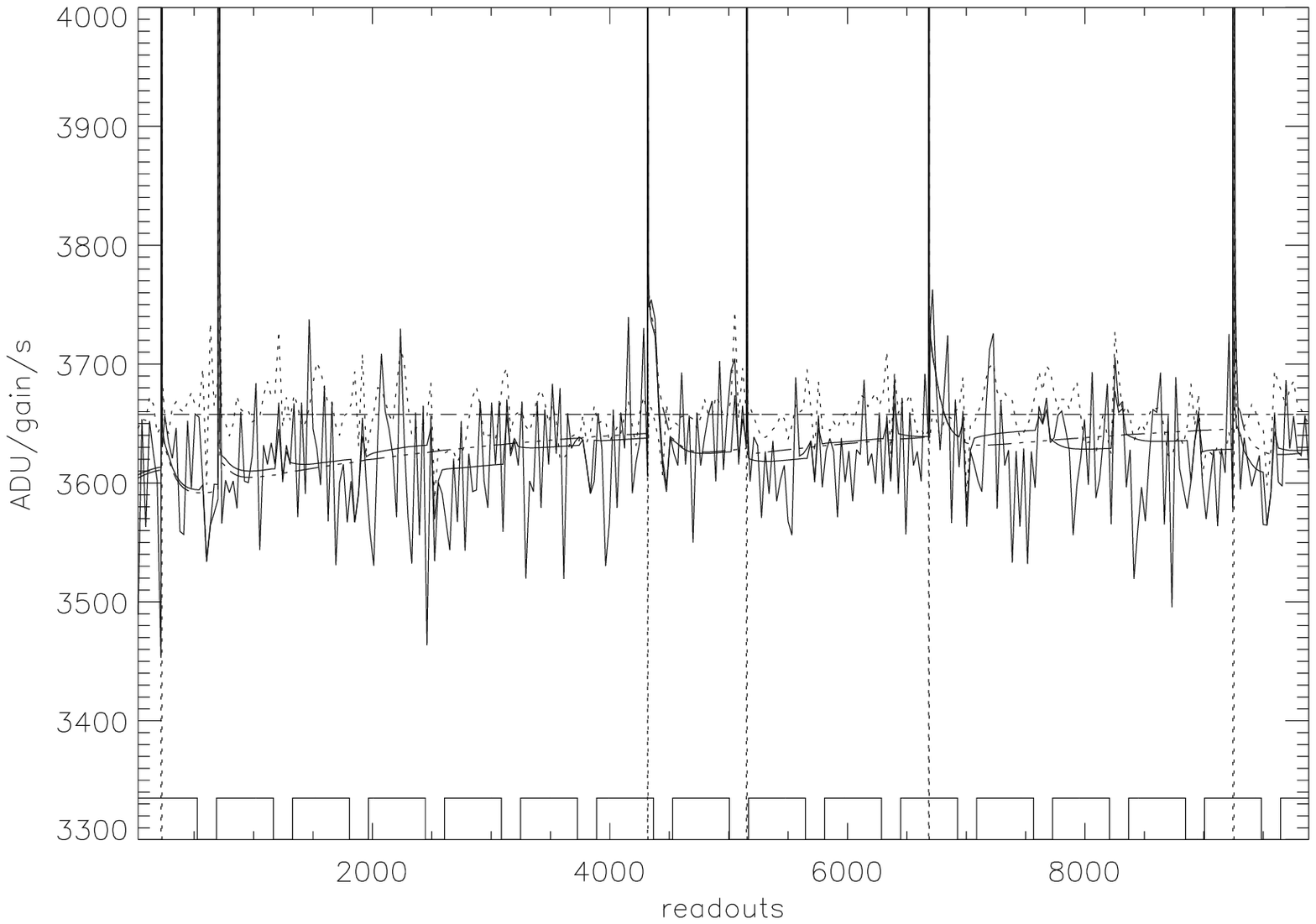, width=14.5cm} \\
2(a)\\
     \epsfig{file=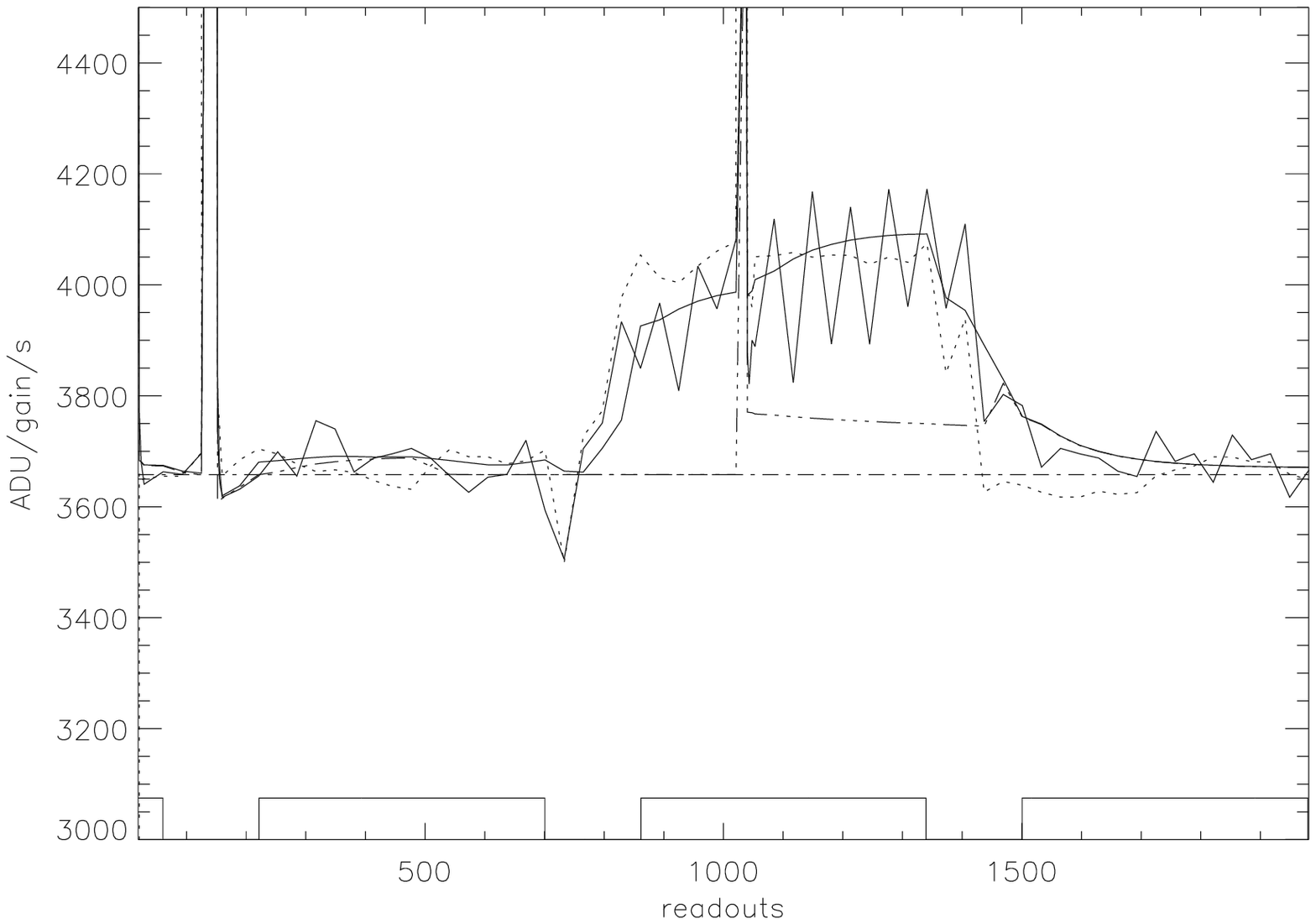, width=14.5cm} \\
2(b)\\
  \end{center} 
\caption{
The Figure shows an Example of real and model data throughout the pixel history.
Top panel (2a): the solid ``noisy'' line represents 
the observed data, with superimposed the best-fit model (continous line), 
which follows the transient trends due to the presence of 
cosmic rays. The dotted line is the data corrected for 
transients and deglitched ({\bf reconstructed} signal). 
The dot-dashed horizontal line is the assumed stabilization background,
while the local background corresponds to the three-dots dashed line. 
In the bottom panel (2b) it is shown how the code works when it sees a bright source. 
In this case the real data have been smoothed. 
} 
\label{model} 
\end{figure*} 
 
With the previously defined quantities we can define two different kinds of fluxes: 
 
1) ''\textbf{Unreconstructed}'' fluxes and maps are derived from the 
excess of the measured signal with respect to the `local background', 
and represent the flux excess not recovered from transients but only 
from glitches. 
 
2) ''\textbf{Reconstructed}'' fluxes and maps are computed 
from the ``reconstruted'' signal, and take into account 
not only the glitches, but also the transients ``on'' the sources. 
 
In other words, the main difference between these two flux estimates appears evident when a pixel 'sees' a source.  
Reconstructed fluxes describe sources taking into account the real excess of signal with respect to the local background plus the transient modelling. The latter 'recovers' for the flux loss due to the slow response of the pixel when it detects an intense prolonged signal, like a source, during a pointing exposure. 
   
Unreconstructed fluxes do not take into account the effects of this source modelling, and thus represent the effective flux collected by the detector during the raster exposure. 
We will see with simulations that our code is not able to properly recover faint sources, and for this reason we will use unreconstructed fluxes in order to generate our final maps.

As an example, Figure \ref{model} shows how the code fits and describes  
the background and the transients. 
In the top panel we plot an example of real and model data 
through the pixel history. The solid ``noisy'' line represents 
the observed data, with superimposed the best-fit model (continous line), 
which nicely follows the transients induced by 
cosmic rays. The dotted line is the data corrected for 
transients and deglitched ({\bf reconstructed} signal). 
The dot-dashed horizontal line is the assumed stabilization background,
while the local background corresponds to the three-dots dashed line. 
The bottom panel shows how the code works when it sees a bright source. 
In this case the real data have been smoothed. 
The `{\bf unreconstructed}' signal is computed as the difference of
the observed data and the local background.

The fitting algorithm starts with the brightest glitches  
identified in the pixel time history, 
assumes discontinuities at these positions, and tries  
to find a fit to the time history that satisfies the mathematical model 
assumed to describe the solid-state physics of the detector. 
In the fit we use the same default parameter values for all the pixels 
(the physical parameters scaled only for the stabilization background), 
leaving as free parameters only the charge values at the beginning of  
the observations and at the 'peaks' of glitches. 
   
By successive iterations, the parameters and the background for each  
pixel are adjusted to better fit the data, until the rms of the difference  
between model and real data is smaller than a given amount  
(e.g. 15 ADU/gain/s ).

\begin{figure} 
  \begin{center} 
    \epsfig{file=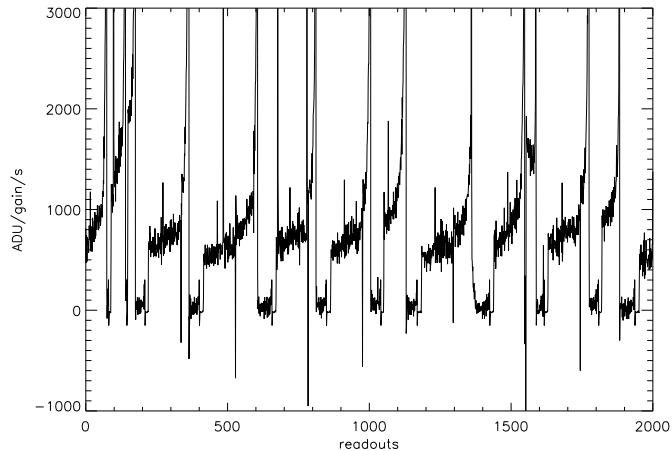, width=9cm} 
\caption{
A repeated series of strong impacts on the detector, that 
can induce the presence of many successive drop-outs. 
Along the pixel time history, the glitches appear as short 
duration high jumps (almost vertical lines) and the slow gradients 
represent the sky signal. Drop-outs affect the sky signal by introducing
discontinuities and responsivity variations. 
The time interval reported in the figure (2000 readouts) 
corresponds to 62.5 seconds.
}

\label{drop_drop} 
  \end{center} 
\end{figure}

All the features described in Figure \ref{model}  
are present in the ISOCAM LW observations as well,  
but sometimes (and not so rarely) we find some peculiar behaviour 
in our data that does not correspond to any usual transient. 
These {\it drop-outs} can be explained in terms of  
saturation of the detector: when a very strong glitch  
impacts on a pixel, this can reach the saturation 
regime (the top level of the instrument dynamical range) 
and then be reset. After the impact, the detector needs 
several readouts before loosing the memory of such a shocking event. 
Usually these {\it drop-outs} appear as isolated features. 
However we found that an intense repeated series  
of impacts can cause much more serious problems to the detector 
electronics, and many {\it drop-outs} can appear consequently 
in the data for a significant fraction of the pixel time history, 
as shown in Figure \ref{drop_drop}. 
It is useful to note here that such long-duration  
{\it drop-outs} are not resolved in ISOCAM, where they 
affect only one or two readouts, because 
of the different temporal resolution (1/32 seconds for PHOT 
against 2 seconds for ISOCAM, for each readout). 
All the readouts affected by this problem must be masked and 
excluded in the fitting procedure. 
 
\subsection{The Interactive Analysis} \label{red2} 

After the first run of the automatic fitting procedure, the next step is  
the interactive mending of fitting failures. 
This massive work of interactive analysis is carried out with an easy-to-use 
widget interface, which allows any kind of repairing which may be necessary. 
This stage strongly depends on the assumed level of the reduction, related 
to the ``goodness'' of the observational data set. For this reason  
observations characterized by different observing parameters (exposure time,  
raster step ...) need a specific treatment during the interactive  
reduction procedure.  
In the case of PHOT we found more efficient to use smoothed data as input for 
the fitting procedure. For the Lockman Hole we chose a smoothing 
factor of 32, implying that the code reads and uses only 1 readout every 
32 (1 second) to which it associates the median value of the previous 32 
readouts. 
Thus short-term features and noise are reduced, allowing a better (and faster) 
modelling of the general trend of the signal.

The automatic detection of sources is checked through ``eye-balling''  
on the time history of each pixel.  
If the code fails (because it finds a source where  
a source is not present, or when it fits improperly a real source), 
a local interactive fit is carried out.  
Furthermore, the limited number of pixels of the PHOT-C100 detector (a  
3x3 array) allows to carefully check the presence of sources scanning the 
time history of each pixel where the median level of the signal in a pointing 
position peaks with respect to nearby values. 
In order to minimize the loss of sources, we also check all the  
raster positions with a signal greater than a given threshold level  
(15 ADU/gain/s for the Lockman Hole).

\subsection{Map generation and source detection} \label{map} 

Once a satisfying fit is obtained for all the pixels over the whole pixel 
history, our pipeline proceeds with the generation of sky maps. An image for 
each raster position is created, by averaging the signals of all the  
readouts relative to that pointing for each pixel. The signal is then 
converted to flux units (mJy pixel$^{-1}$, see Appendix \ref{calib}  
and Section \ref{flux1}), glitches and bad data are masked 
and the images are then combined to create the final raster maps (one for  
each raster position). These images are projected onto a sky map 
(raster image) using the projection algorithm available for 
ISOCAM data in the CIA package (Cam Interactive Analysis, Ott et al., 2001). 
When projecting the signal on the sky, we make use of the nominal 
raster astrometry.

The redundancy of Lockman ISOPHOT observations allowed us to  
generate high resolution maps, rebinning the original data into a  
final map with pixel size of 15$\times$15 arcsec.
The detector signal is distributed in a uniform way between the
smaller pixels. 
This process allows a better determination of source positions. 
 
The source detection is performed on the signal-to-noise maps, given
by the ratio of the ``unreconstructed'' flux maps and the corresponding
maps of the noise. For the source detection we do not need any calibrated map,
a relative map is sufficient in order to find the positions of any 
positive brightness fluctuation (as discussed by Dole et al., 2001).

First, our task selects all pixels above a low flux 
threshold (0.6 mJy pixel$^{-1}$) using the 
IDL Astronomy Users Library task called $find$ (based on DAOPHOT's 
equivalent algorithm). This algorithm finds positive 
brightness perturbations in an image, returning centroids and 
shape parameters (roundness and sharpness).
The algorithm has been carefully tuned in order to detected all sources
and to miss only the spurious ones that could be found surrounding 
the brighest sources or close to the edges of the maps
(the input parameters are in particular the FWHM of the instrument
and the limits for the roundness and sharpness geometric acceptable
galaxy values).

Then we extract from the 
selected list only those objects having a signal-to-noise ratio $>$ 3. 
 
In the final stage of our reduction we use our simulation procedures to 
reproject the sources detected on the raster map onto the pixel time history 
(see Section \ref{simul}). 
In this way we are able to check the different temporal positions supposed  
to contribute to the total flux of each source. This method allows to improve  
the fit of the data for all the sources that appear above the interactive  
check threshold, and to significantly reduce the flux defect of the  
detected sources.

\section{SIMULATIONS} \label{simul} 

The only way to assess the capability of our data reduction method for
source detection and flux estimate is through simulations. 
We use the ISOPHOT C100 PSF (stored in the PIA file PC1FOOTP.FITS),  
rebinned in order to have the same pixel size (15''x15'') of the final maps  
in which the source extraction has been performed.  
Through the projection task used to produce maps, we project  
the PSF scaled to any given input flux on the raster maps. 
This corresponds to generate a synthetic source in the real map. 
Furthermore, our code is able to modify the time histories in  
those raster positions (of every pixel) contributing to the  
total flux of a given source. This is done by converting the given 
input flux excess (in respect to the local background)  
from mJy to ADU/gain/s, and adding it to the real pixel histories 
(containing glitches and noise). When this excess is added  
to the underlying signal, it is modelled by the algorithm that takes 
into account the transient response of the detector pixels (see 
Appendix \ref{simul_app} for details) and finally reproduces a real source, 
as it usually appears along any original pixel time history.  
 
In our approach we make use of this powerful instrument at two  
different but complementary levels: 
 
1- {\it Simulation of detected sources in the same positions as they are  
detected on raster maps}.  
As already anticipated in Section \ref{map}, with this procedure 
we can improve the data reduction and the reliability of our final  
sample. Simulating all detected sources with signal-to-noise 
ratio $>$3, we can check  
all the pieces of pixel histories supposed to contribute to them. 
In this way we can reject all spurious detections, and perform 
a better fit when needed.   
 
2- {\it Simulation of a sample of randomly located sources}, 
in order to study the completeness and reliability of our detections 
at different flux levels, and estimate the internal calibration of the  
source photometry. 
The strategy we have adopted for the Lockman Hole is here described. 
We added 40 randomly distributed point sources at five different  
total fluxes (37, 75, 150, 300, 600 mJy).  
To avoid confusion, we imposed a minimum distance 
of 135 arcsec ($\sim$ 3 pixels) between the simulated and real 
sources on the maps. For the same reason, at each flux level  
we have distributed the 40 sources between the four rasters  
(each covering an area of $\sim$ 0.13 deg$^2$): 10 sources  
per raster, divided in 2 simulation runs each containing 5 sources.

\begin{figure} 
  \begin{center} 
    \epsfig{file=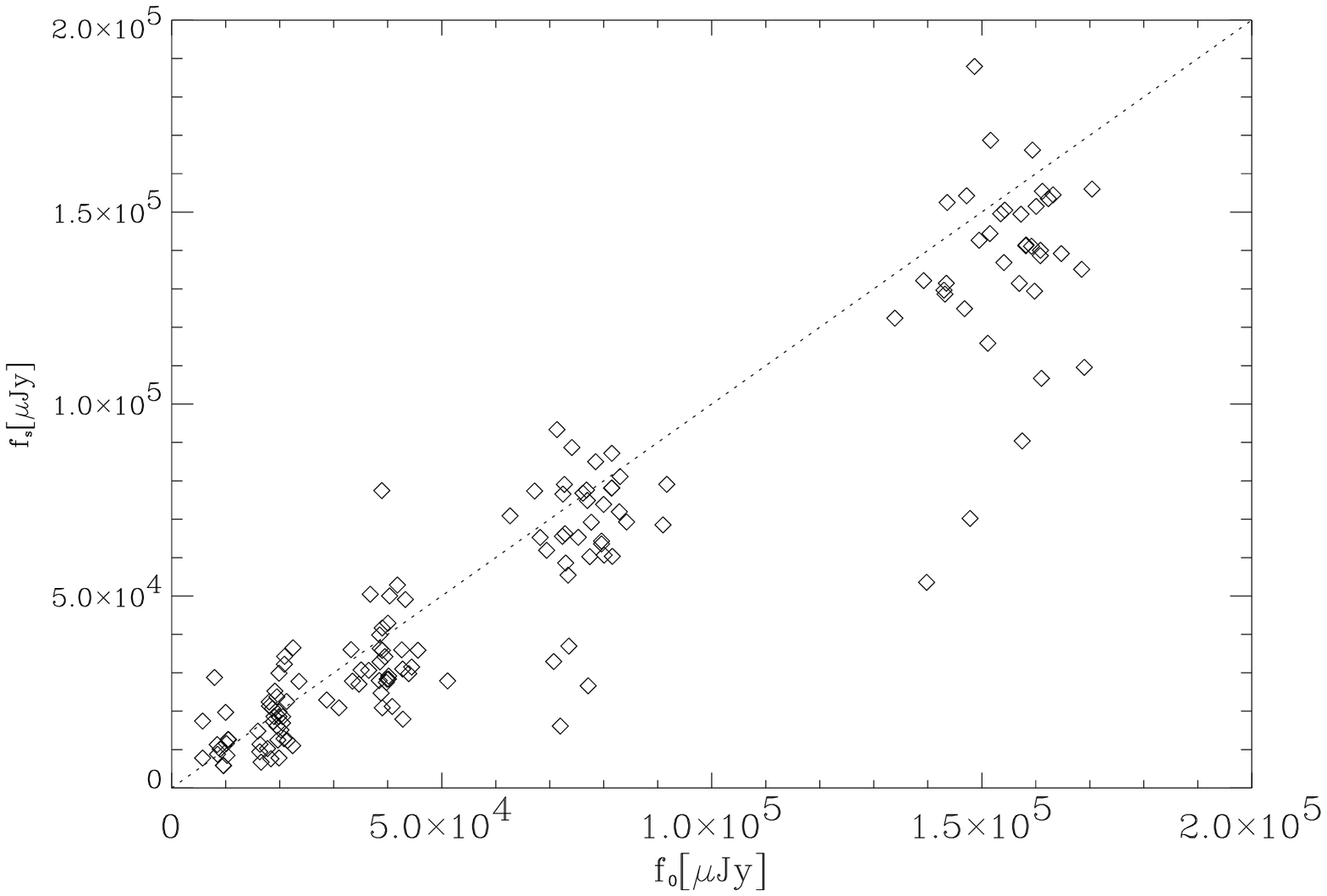, width=9cm} 
    \epsfig{file=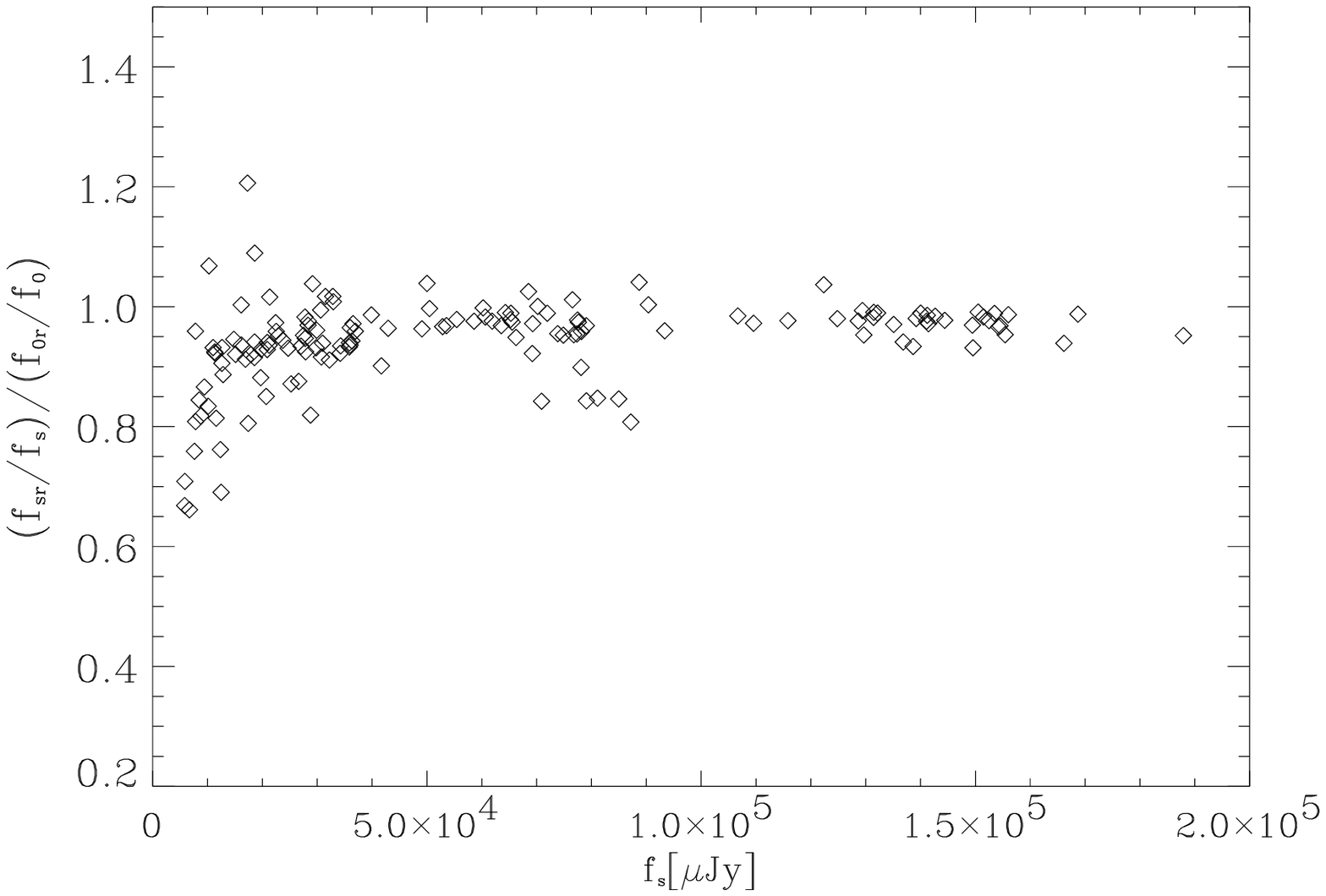 , width=9cm} 
  \end{center} 
\caption{{\it top} - Output peak flux after reduction versus output peak  
flux due only to sampling effects for simulated sources. The fluxes are  
{\bf `unreconstructed'} for transients.   
{\it bottom} - {\bf `Reconstructed'}-to-{\bf `unreconstructed'} flux  
ratio ($fsr/fs$) normalized to the same ratio obtained for mapping  
effects only ($f0r/f0$) as function of the detected {\bf `unreconstructed'} 
peak flux. The ratio distribution broadens towards lower values at  
faint fluxes, due to the characteristic of our reduction method of  
not reconstructing faint sources.} 
\label{f0_fs} 
\end{figure} 
 
\begin{figure} 
  \begin{center} 
    \epsfig{file=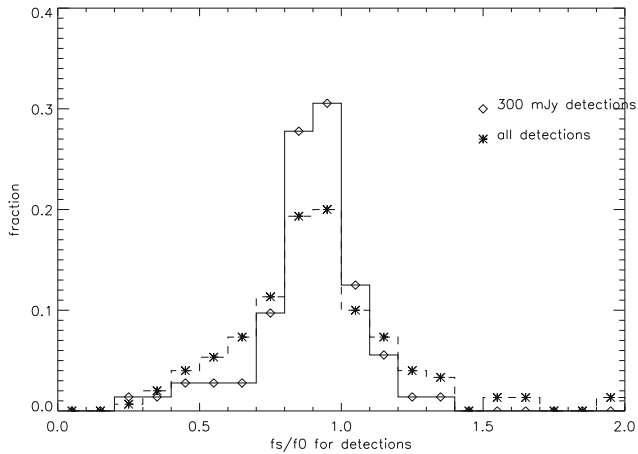, width=9cm} 
  \end{center} 
\caption{Distribution of the ratio between the reduced peak flux and 
the unreduced peak flux (derived from mapping effects only), $fs/f0$,  
for all the simulated sources detected above $3 \sigma$ (dashed  
line marked by diagonal crosses) and for the   
detections injected above 300 mJy (solid line marked by diamonds).} 
\label{fs_f0_dist} 
\end{figure} 
 
We reduced the simulated data cubes exactly in the same way as we did  
for the original data, doing the same checks and repairs.  
We produced the simulated maps on which we extracted the simulated sources, 
following the procedures used for real rasters. For each detected simulated source 
we have measured positions and peak fluxes.  
As in L01, the peak fluxes measured on the maps will be referred to  
as $f_s$ and $f_{sr}$ (both for real and synthetic sources) respectively  
for {\bf `unreconstructed'} and {\bf `reconstructed'} 
maps. The corresponding theoretical peak fluxes associated to the  
excess flux maps, not reduced and containing neither glitches or  
noise, will be named $f_0$ and $f_{0r}$.  
The theoretical quantities are produced only by the Lockman Hole 
observational strategy and the ISOPHOT instrument, while the measured 
quantities are also affected by our reduction method. 
These simulated data cubes contain both real sources 
and simulated ones. They have also the same noise, the `glitches',  
and background transients as the original data.  
 
In figure \ref{f0_fs} ({\it top}) we compare the output peak fluxes obtained  
for the simulated sources affected only by the mapping effects ($f_0$) with the output fluxes of the reduced simulated sources ($f_s$). 
There is a linear correlation, and we observe that the reduced fluxes are always  
slightly lower than the unreduced ones.  
We find a similar correlation in the corresponding 
{\bf `reconstructed'} peak fluxes ($f_{0r}$ and $f_{sr}$), although for faint 
sources our algorithm is not able to reconstruct correctly the fluxes (see 
Figure \ref{f0_fs}, $bottom$).  
In order to have a correct flux determination at each level (both bright and faint), 
we will always use the {\bf `unreconstructed'} fluxes.
 
Figure \ref{fs_f0_dist} shows the distribution of the ratio between $f_s$ and $f_0$, 
for the simulated sources detected above $3 \sigma$, compared to the same ratio for 
the detections above 300 mJy only.  
The peak of the distribution at 0.88 indicates a general underestimation of the total 
fluxes (derived from $f_s$).  
Two combined effects can explain this failure of the code to correctly compute 
total fluxes: one is the fact 
that $f_0$ is computed on the measured positions of $f_s$;  
the other is the underestimate of the wings of faint sources by  
the reduction method. 
The predicted and unbiased distribution of $f_s/f_0$ for 
the brighter simulated sources (above 300 mJy) peaks at 0.88.  
This value was subsequently assumed to correct our measured fluxes.

\begin{figure} 
  \begin{center} 
    \epsfig{file=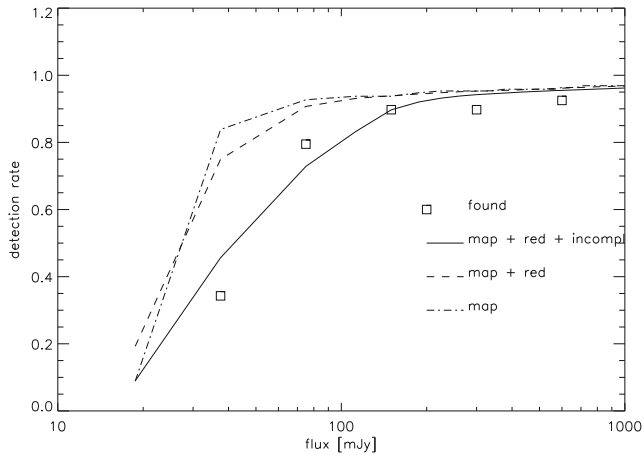, width=9cm} 
  \end{center} 
\caption{Percentage of detected sources in simulations of the Lockman Hole  
as a function of the source input flux. The detection threshold   
has been set to 3$\sigma$. The dash-dotted curve  
represents the effect of PSF sampling on the detection rate; the dashed curve  
represents the effect of the reduction; the solid curve is the total effect on 
reduced PHOT C100 Lockman data.} 
\label{compl_fig} 
\end{figure}

\begin{table} 
\caption{Detection rate in the simulations.} 
\label{det_rate} 
\begin{tabular}{|c|c|c|c|} 
\hline 
\hline 
Input~Flux~(mJy)&Number of &Number of&Detection rate in\\
~&simulated sources&detected sources&the simulations\\
\hline 
\hline     
37  & 40 & 12 & 30.0~\% \\ 
75  & 40 & 31 & 77.5~\% \\ 
150 & 40 & 35 & 87.5~\% \\ 
300 & 40 & 35 & 87.5~\% \\ 
600 & 40 & 37 & 92.5~\% \\                                                      
\hline 
\hline 
\end{tabular} 
\end{table}

\subsection{Completeness} \label{compl} 

As mentioned in the previous section, with simulations in the Lockman Hole 
we have derived the distribution of the measured ($f_s$) to theoretical ($f_0$)  
peak flux ratio. This distribution is crucial in deriving the completeness  
of the catalogue and the internal flux calibration, as it allows to predict the number
of detected sources at a given flux level (the  
$g-function$ described in detail in Gruppioni et al. 2002).  
 
Our data reduction method can introduce some additional incompleteness if a source 
is interpreted as a background transient, and lost from the final source catalogue. 
We can estimate the incompleteness of our method from simulations, by computing the ratio between the number of detections and the number of expected sources in different peak flux intervals.  
 
\begin{figure} 
  \begin{center} 
    \epsfig{file=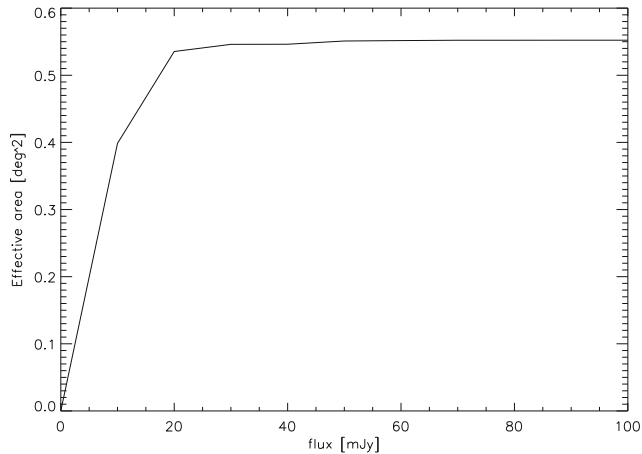, width=9cm} 
  \end{center} 
\caption{Effective area versus total flux density for the 95 
$\mu$m catalogues in the Lockman Hole area.} 
\label{area} 
\end{figure}

The results of the simulations are reported in Table \ref{det_rate} 
and shown in Figure \ref{compl_fig}, where the resulting function  
describing the incompleteness of our survey is plotted as a 
function of the simulated input flux. 
The loss of bright sources happens only when they are located 
at the extreme edges of the raster maps. 
We need to consider also the areal coverage of our survey  
(i.e. the fraction of the survey area where a source of peak flux  
$f_s$ can be detected with $S/N > 3\sigma$), which is reported in 
Figure \ref{area}.   

The global correction to be applied to the observed source counts
is then obtained by convolving the function describing the completeness 
of our method with the areal coverage function 
(see L01 and Gruppioni et al. 2002 for details).

\section{FLUX DETERMINATION} \label{flux} 

The simulations performed in the Lockman field provided not only the  
completeness of our detections at different flux levels, but also the  
internal calibration of the source photometry and the distribution of  
the ratio between the measured and the theoretical peak fluxes. 
 
Here we summarize some relevant definitions and the relations used 
to derive the final total fluxes of real detected sources, as in L01 
and Gruppioni et al. (2002): 
 
\begin{itemize} 
\item $f_s$: is the peak flux measured on maps for both real and  
simulated sources. 
Its value depends both on data reduction method, on the Lockman  
observing strategy, and the ISOPHOT instrumental effects; 
\item $f_0$: is the `theoretical' peak flux measured on simulated  
maps containing neither glitches or noise. Its value depends only on  
Lockman observing strategy and the ISOPHOT instrumental effects; 

\item $q \equiv f_s / f_0$; 

\item $q_{med}$: is the peak of the $f_s / f_0$ distribution (also called 
systematic flux bias) and is 0.88. 
This value is used to correct the measured flux densities; 
 
\item the flux density $s$ of a source is computed by applying a correction 
factor to the measured peak flux $f_s$ in order to have a measure 
of its 'total' flux. This can be done using the output informations 
stored after the simulation process of real sources in the positions where 
they have been detected (see Section \ref{simul}). This gives the ratio 
between the injected simulated total flux $s_0$ and the corresponding theoretical peak flux $f_0$ measured on the simulated maps. 
This ratio represents the correction needed to derive total fluxes 
from peak fluxes. In this way the flux density is:  
\begin{equation}  
s = \left(f_s \times \frac{s_0}{f_0}\right) / q_{med} \label{ftot} 
\end{equation} 
 
For simulations $s_0$ is the injected total flux, while for real data 
$s_0$ is derived through successive iterations starting from a rough estimated 
value:  
\begin{equation} 
s_0 = f_s / \left<f_s/s_0\right>_{sim} \label{flux_eq} 
\end{equation}                                           
where $\left<f_s/s_0\right>_{sim} = 0.23$ is the average value resulting from  
simulating a point source of unitary flux.
We can consider $s$ as the measured flux density and $s_0$ as the `true'  
flux density of a source; 

\item finally we need to increase the derived flux density of a factor 1.2. 
This correction is needed because the PSF we use to simulate sources is  
undersampled and it misses the flux presents in the external wings of an 
ISOPHOT source profile.  
\end{itemize}

\subsection{Calibration} 
\label{flux1} 
\begin{table} 
\caption{
Comparison of fluxes obtained with our reduction against  
independent estimates for a sample of sources. In case of IRAS 
detection, the comparison fluxes have been derived from interpolation 
between two IRAS bands (60 and 100 $\mu$m). 
For stars without IRAS counterpart, we report SED model predictions.
We also report for each source the reference for the comparison flux
(GBPP = ISO Ground Based Preparatory Programme, PSC = IRAS Point
Source Catalogue, see text for details). 
} 
\begin{tabular}{|l|l|l|l|l|l|} 
\hline 
\hline 
Target&Our&Comparison&Ref.&S/G&comparison\\ 
~&Flux (mJy)&Flux (mJy)&~&~&source\\ 
\hline 
\hline                                                           
HR5981      &   67 $\pm$13&   77$\pm$3&   (GBPP)    &  star   &  SED model        \\ 
HR6464      &   181$\pm$32&  148$\pm$5&   (GBPP)    &  star   &  SED model        \\ 
HR6132      &   343$\pm$66&  348$\pm$10&  (GBPP)    &  star   &  SED model        \\ 
HR1654      &   913$\pm$200&  712$\pm$70& (IRAS PSC)&  star   &  IRAS       \\ 
F16344+4111 &   655$\pm$150&  648$\pm$65 &(IRAS PSC)&  galaxy &  IRAS       \\ 
F10507+5723 &   577$\pm$110& 1000$\pm$100&(IRAS PSC)&  galaxy &  IRAS       \\ 
\hline 
\hline 
\end{tabular} 
\label{tab_cal} 
\end{table}

We have computed a new estimate of the detector responsivities, 
that converts digital units [ADU/gain/s] to physical flux units [mJy]. 
The detailed description is reported in Appendix \ref{calib}.  
In order to check the general consistency of our calibration,  
we have reduced a set of external calibrators with our procedures, 
by following the same steps described in the previous sections. 

It is very difficult to find good ``standards'' for far-infrared 
observations. An absolute calibration is hampered by the intrinsic  
uncertainties and by the low sensitivities of previous  
measurements (IRAS, COBE-DIRBE), especially at faint fluxes. 
Our set of ``calibrators'' includes four stars and two IRAS sources,  
as reported in Table \ref{tab_cal}. 
For IRAS sources we have a direct measure of the far-infrared flux 
(from the IRAS Point Source Catalogue). 
To get the 95 $\mu$m flux we made an interpolation between the 
fluxes at 60 and 100 $\mu$m. 
For stars we compare our fluxes with predictions from 
Spectral Energy Distribution (SED) models. In particular we select stars 
from the ISO Ground-Based Preparatory Programme (GBPP, Jourdain de Muizon, M. 
and  Habing, H. J., 1992). Synthetic SEDs  
for these stars are available in the ISO calibration Web page  
($http://www.iso.vilspa.esa.es/users/expl\_lib/ISO/wwwcal/$). 
The optical and near-infrared photometry observed in the GBPP  
have been fitted with a Kurucz model to provide the flux densities at  
longer wavelengths, thus extending the SED out to 300 $\mu$m (Cohen et al., 
1996). 
We have chosen to adopt these SEDs model fluxes for use in our calibration 
at 95 $\mu$m, after convolution with the PHOT C\_95 filter. 
A summary of the fluxes obtained with our reduction and the expected 
fluxes are reported in Table \ref{tab_cal}.  
Quite good agreement with the comparison fluxes is found over a wide range of fluxes. 
Only for source F10507+5723, the only IRAS galaxy in the Lockman Hole, 
our value is a factor 2 lower than the interpolated IRAS flux. 
Given the uncertainties of the IRAS 100 $\mu$m flux ($\sim 10-20$\%), 
we have chosen to assume our calibration as final estimate of fluxes, without  
applying any other scaling factor. 
Excluding this IRAS source, Table \ref{tab_cal} shows that, over a wide range  
of fluxes, the error on the photometry is within 20\%. 
A more secure constraint on the calibration will be available after 
the reduction of ELAIS 95 $\mu$m fields ($\sim 15$ deg$^2$) that 
will provide a wider sample of IRAS galaxies (few tens).

\subsection{Flux errors} 
 
To compute the errors associated to our flux estimates, we have taken
into account the two major contributions to the uncertainties.
The first one depends on the data reduction method, given by the distribution
of the ratio between fluxes measured after the reduction ($f_s$), and
those measured taking into account only mapping effects ($f_0$).
If $f_0$ is the real flux
and if we measure after the data reduction $f_s$, it means that the reduction
has introduced an error in the measure (by modifing the real flux).
We can estimate this error as the width of the distribution of $f_s/f_0$.
The second error term is due to the noise of the map.

Combining these two quantities, we get the final errors on the photometry
(as reported in Table \ref{catalog}). The median of the errors distribution
is $\sim20$\%.

\subsection{Positional accuracy} 

With the set of simulations used to derive the completeness and the 
photometric corrections, we can also estimate the uncertainties on the 
astrometric positions. 
For each simulated source we have the injected and the measured coordinates. 
The local background noise can affect the estimate 
of the position of each galaxy centroid. The resulting effect is that a source, 
simulated in a given position, will be detected by the extraction algorithm 
in a slightly but different location. 

\begin{figure} 
  \begin{center} 
    \epsfig{file=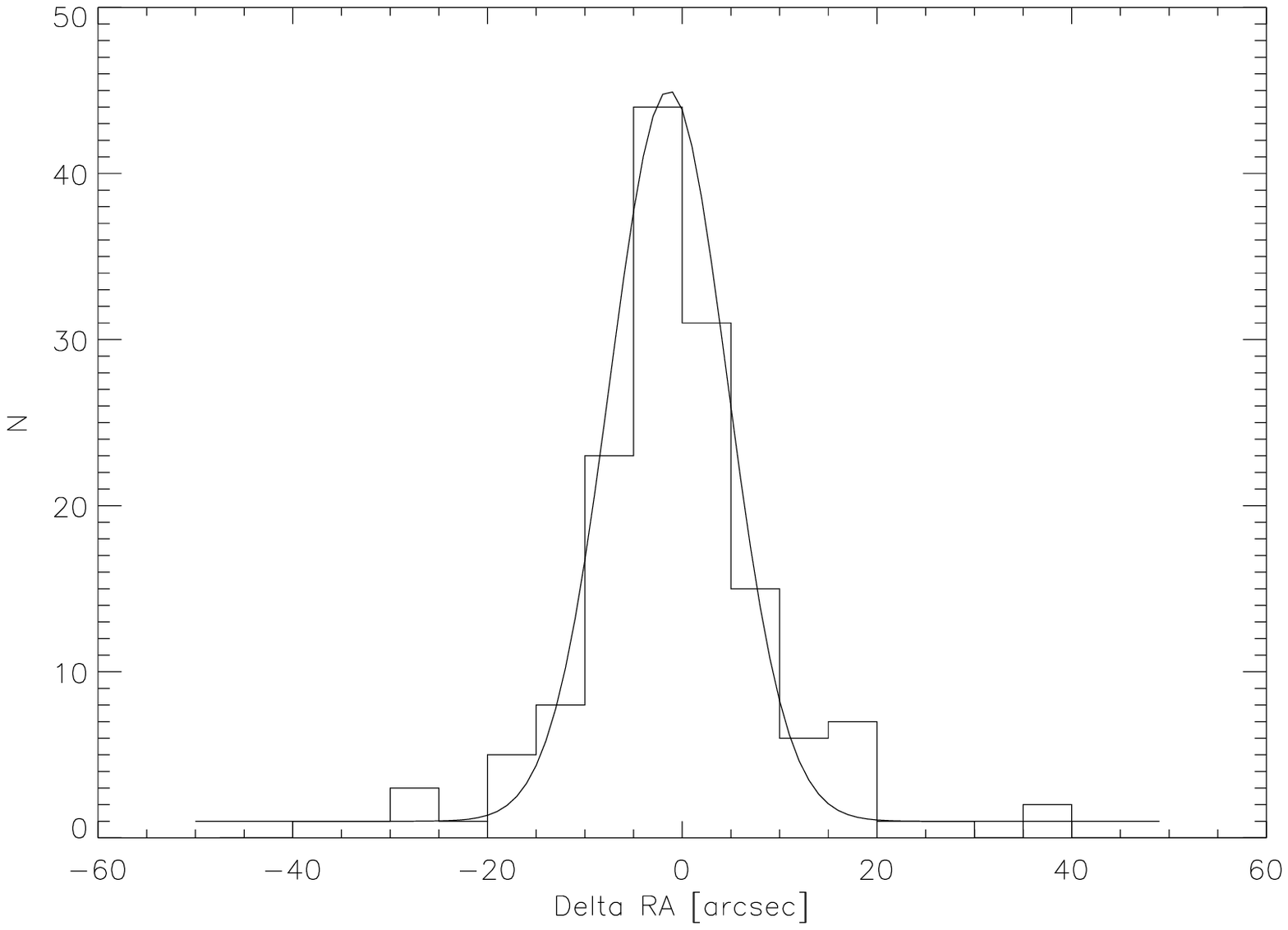 , width=9cm} 
    \epsfig{file=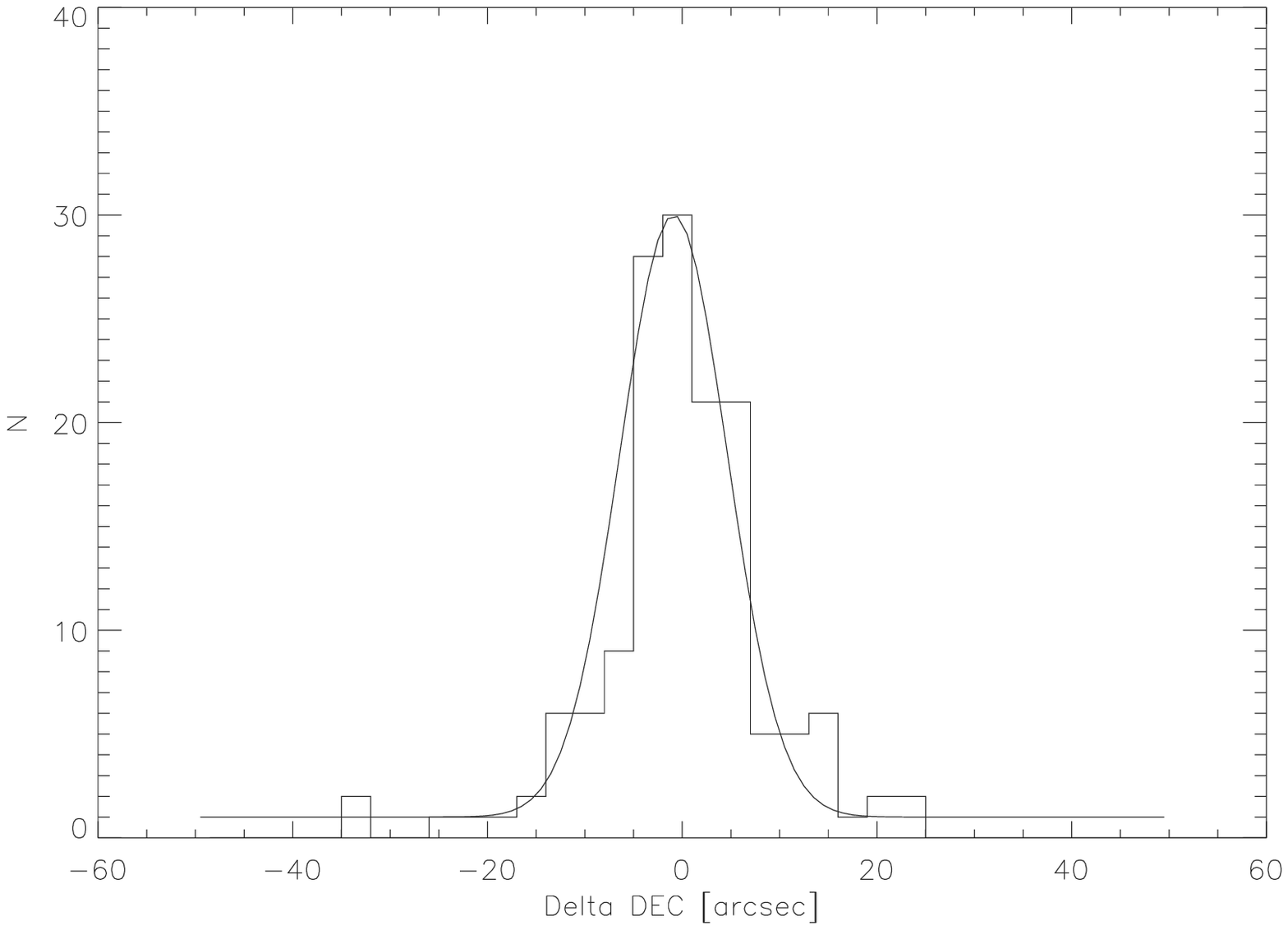 , width=9cm} 
  \end{center} 
\caption{Distribution of the difference in RA ({\it top}) and  
DEC ({\it bottom}) between the injected and the found positions for  
the simulated sources.} 
\label{position} 
\end{figure}

Figure \ref{position} shows the distribution of the differences in RA 
({\it top}) and DEC ({\it bottom}) between the injected and the estimated 
positions for the simulated sources. 
The mean positional accuracy is of the order of 20 arcsec. 
By fitting the histograms in Fig. \ref{position}
with a Gaussian function, the 3-$\sigma$ of the distributions are
respectively 18 arcsec for the right ascension and 17 arcsec for the declination.
Of the 150 simulated and detected sources, 90 (60\%) lie within
[-10,+10] arcsec.

\section{OBSERVATIONS OF AN AREA IN THE LOCKMAN HOLE}

The 95 $\mu$m observations of the Lockman Hole field (P.I. Y. Taniguchi)
by the photo-polarimeter ISOPHOT represent one of the best dataset available
in the ISO archive to test the performance of our reduction technique. 
The long elementary integration times ($\sim 16$ sec for each raster position)
and the observing redundancy allow us an accurate evaluation  
and modelling of any transient effects, for both long and short time scales.

\subsection{The ISOPHOT observation strategy}  \label{observation} 
\begin{figure*} 
  \begin{center} 
    \epsfig{file=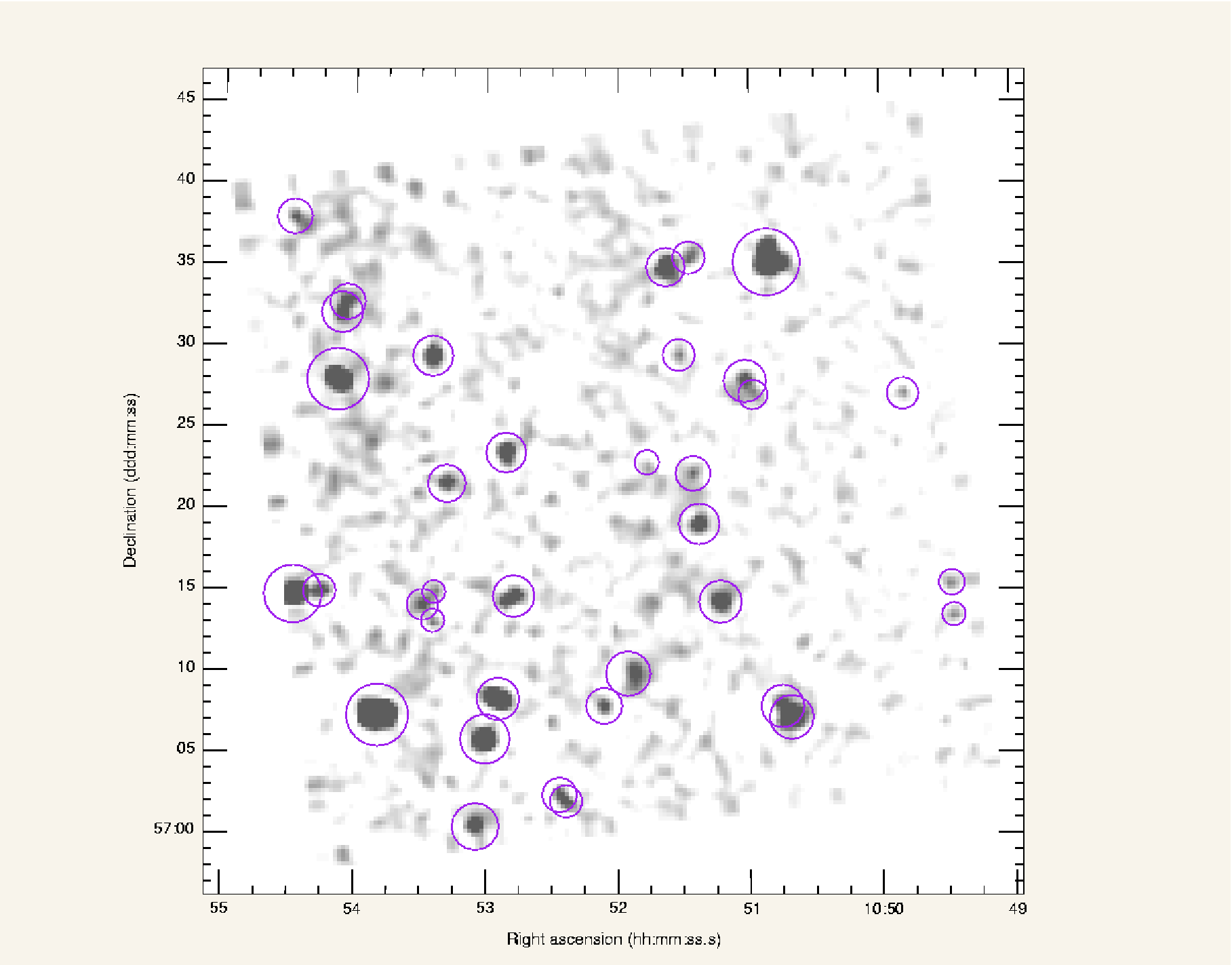, width=18cm} 
 \end{center} 
\caption{The 95 $\mu$m map of the Lockman Hole LHEX area. The circles indicate 
our detected sources, with a size proportional to the source flux.} 
\label{mosaic} 
\end{figure*} 

The Lockman Hole (Lockman et al., 1986) was selected for its high  
ecliptic latitude ($|\beta| > 50$), to keep the Zodiacal dust emission 
at the minimum, and for the low cirrus emission. This region presents the lowest  
HI column density in the sky, hence is particularly suited for the  
detection of faint infrared extragalactic sources. This consideration triggered a number
of multifrequency observing campaigns in the past several years.
The spectral coverage includes the X-rays (e.g. Hasinger et al., 2001), 
the optical (e.g. Fadda et al. 2002, in preparation), the  
mid-infrared (Fadda et al., 2002), the far-infrared (Kawara et al. 1998, and
this work), the submillimiter (Scott et al., 2002), and the radio bands
(De Ruiter et al., 1997; Ciliegi et al. 2003, in preparation).  
The published spectroscopic information is on the contrary still sparse. 
 
Two different regions in the Lockman Hole have been observed by ISOPHOT.  Each one of  
the two fields, called LHEX and LHNW, covers an area of  $\sim$ 44'x44' and has been 
surveyed at two far-infrared wavelengths with the C100 and 
C200  detector (respectively at 95 and 175 $\mu$m, with the C\_90 and C\_160  
filters), in the P22 survey raster mode (see Kawara et al., 1998). 
We focused our analysis on the LHEX field, which consists in a mosaic    
of four rasters, each one covering an area of $\sim$ 22'x22'. 
The ISOPHOT detector was moved across the sky describing a grid pattern, with about 
half detector steps (corresponding to 1.5 detector pixel, or 67 arcsec) in both directions. 
This strategy improves the reliability of source detections and the image quality,
as each sky position is observed twice in successive pointings. 
Table \ref{lockman} summarizes the observational parameters for the C\_90 filter. 
These data have been retrieved from the ISO data archive through the WEB
interface $http://isowww.estec.esa.nl/$. 
 
\begin{table} 
\caption{Lockman Hole ISOPHOT 95$\mu$m observational parameters.} 
\begin{tabular}{|l|l|} 
\hline 
\hline 
integration time per sample             & $1/32~sec$\\ 
integration time per pointing           & $16~sec$\\ 
total integration time per pixel and per raster     & $1.44~hours$\\
number of horizontal and vertical steps  & $18,18$\\ 
step sizes                              & $69'',69''$\\ 
grid size				& $42'x42'$\\
redundancy				& $4$\\
total area covered                      & $0.5~ deg^2$\\ 
equatorial coords of the field center	&RA=10h52m00s~Dec=+57d20m00s\\
galactic coords	of the field center	&Long=149.5deg~Lat=+53.17deg \\
\hline 
\hline 
\end{tabular} 
\label{lockman} 
\end{table}

\subsection{The 95 $\mu$m source catalogue in the Lockman Hole} 
\label{catalog_sec} 

\begin{table*} 
\caption{} 
 
\begin{tabular}{|l|c|c|c|c|c|} 
\hline 
\hline 
ID& RA     &  DEC    & S/N & Flux  \\ 
~ &(J2000) & (J2000) &~    & [mJy] \\
\hline 
\hline   
LHJ105324+572921  &10:53:24.5  &  +57:29:21  &    101 &95   $\pm$  18\\ 
LHJ105250+572325  &10:52:50.9  &  +57:23:25  &    108 &99   $\pm$  19\\ 
LHJ105349+570716  &10:53:49.1  &  +57:07:16  &   59   &577  $\pm$  110\\ 
LHJ105052+573507  &10:50:52.0  &  +57:35:07  &   27   &224  $\pm$  42\\ 
LHJ105407+572753  &10:54:07.9  &  +57:27:53  &   25   &347  $\pm$  66\\ 
LHJ105427+571441  &10:54:27.8  &  +57:14:41  &   24   &281  $\pm$  54\\ 
LHJ105300+570548  &10:53:00.3  &  +57:05:48  &   20   &169  $\pm$  32\\ 
LHJ105041+570708  &10:50:41.2  &  +57:07:08  &   18   &126  $\pm$  24\\ 
LHJ105254+570816  &10:52:54.4  &  +57:08:16  &   17   &145  $\pm$  27\\   
LHJ105138+573448  &10:51:38.2  &  +57:34:48  &   16   &139  $\pm$  26\\ 
LHJ105123+571902  &10:51:23.0  &  +57:19:02  &   10   &90   $\pm$  17\\ 
LHJ105113+571415  &10:51:13.4  &  +57:14:15  &   10   &91   $\pm$  17\\ 
LHJ105155+570950  &10:51:55.3  &  +57:09:50  &    8   &98   $\pm$  19\\ 
LHJ105304+570025  &10:53:04.6  &  +57:00:25  &    8   &86   $\pm$  17\\ 
LHJ105102+572748  &10:51:02.0  &  +57:27:48  &    7   &52   $\pm$  10\\ 
LHJ105406+573201  &10:54:06.1  &  +57:32:01  &    7   &61   $\pm$  12\\ 
LHJ105045+570749  &10:50:45.3  &  +57:07:49  &    7   &25   $\pm$  5\\ 
LHJ105247+571435  &10:52:47.4  &  +57:14:35  &    7   &77   $\pm$  15\\ 
LHJ105318+572130  &10:53:18.0  &  +57:21:30  &    7   &63   $\pm$  12\\ 
LHJ105127+573524  &10:51:27.6  &  +57:35:24  &    6   &49   $\pm$  9\\ 
LHJ105403+573240  &10:54:03.6  &  +57:32:40  &    6   &41   $\pm$  8\\ 
LHJ105223+570159  &10:52:23.5  &  +57:01:59  &    6   &32   $\pm$  6\\ 
LHJ105415+571453  &10:54:15.8  &  +57:14:53  &    6   &44   $\pm$  9\\ 
LHJ105125+572208  &10:51:25.7  &  +57:22:08  &    6   &77   $\pm$  15\\ 
LHJ105206+570751  &10:52:06.2  &  +57:07:51  &    6   &62   $\pm$  12\\ 
LHJ105428+573753  &10:54:28.1  &  +57:37:53  &    5   &70   $\pm$  13\\ 
LHJ105328+571404  &10:53:28.9  &  +57:14:04  &    5   &53   $\pm$  10\\ 
LHJ105226+570222  &10:52:26.4  &  +57:02:22  &    5   &27   $\pm$  5\\  
LHJ105132+572925  &10:51:32.2  &  +57:29:25  &    4   &45   $\pm$  9\\ 
LHJ105058+572658  &10:50:58.4  &  +57:26:58  &    4   &16   $\pm$  3\\ 
LHJ104949+572701  &10:49:49.9  &  +57:27:01  &    4   &67   $\pm$  13\\ 
LHJ104928+571523  &10:49:28.2  &  +57:15:23  &    4   &33   $\pm$  6\\ 
LHJ105146+572249  &10:51:46.8  &  +57:22:49  &    3   &41   $\pm$  8\\  
LHJ104927+571325  &10:49:27.4  &  +57:13:25  &    3   &32   $\pm$  6\\ 
LHJ105324+571305  &10:53:24.3  &  +57:13:05  &    3   &18   $\pm$  4\\ 
LHJ105323+571451  &10:53:23.7  &  +57:14:51  &    3   &20   $\pm$  4\\ 
\hline 
\hline 
\end{tabular} 
\label{catalog} 
\end{table*}

The final catalogue obtained with our method contains 36 sources detected 
at 95 $\mu$m in the Lockman Hole LHEX, over an area of 0.5 deg$^2$. 
All sources have a signal-to-noise ratio greater than 3 and a flux
greater than 16 mJy. 

In Figure \ref{mosaic} we show the final mosaiced map obtained by combining 
together the four rasters. 
The open circles, whose sizes are roughly proportional to
the source fluxes, indicate our detected sources. 
IAU-conformal names, sky coordinates (right ascension and declination at 
Equinox J2000), the detection significance (signal-to-noise ratio), 
the 95 $\mu$m total fluxes (in mJy) 
and their uncertainties are reported in Table \ref{catalog}. 
As previously discussed, all sources have been extracted from the 
map and confirmed by visual inspection on the pixel history (by two 
independent people). This approach produces an highly reliable source
catalogue. 
 
In a forthcoming paper we will discuss the optical, radio, 
mid-IR identifications of the Lockman 95 $\mu$m sources 
with radio and ISOCAM counterparts  
(Rodighiero et al. 2003, in preparation; Fadda et al. 2003, in preparation;
Aussel et al. 2003, in preparation) 
and we will analyse the nature of our far-IR sources and their redshift 
distribution where the spectroscopic information is available.

 \begin{figure*} 
   \begin{center} 
     \epsfig{file=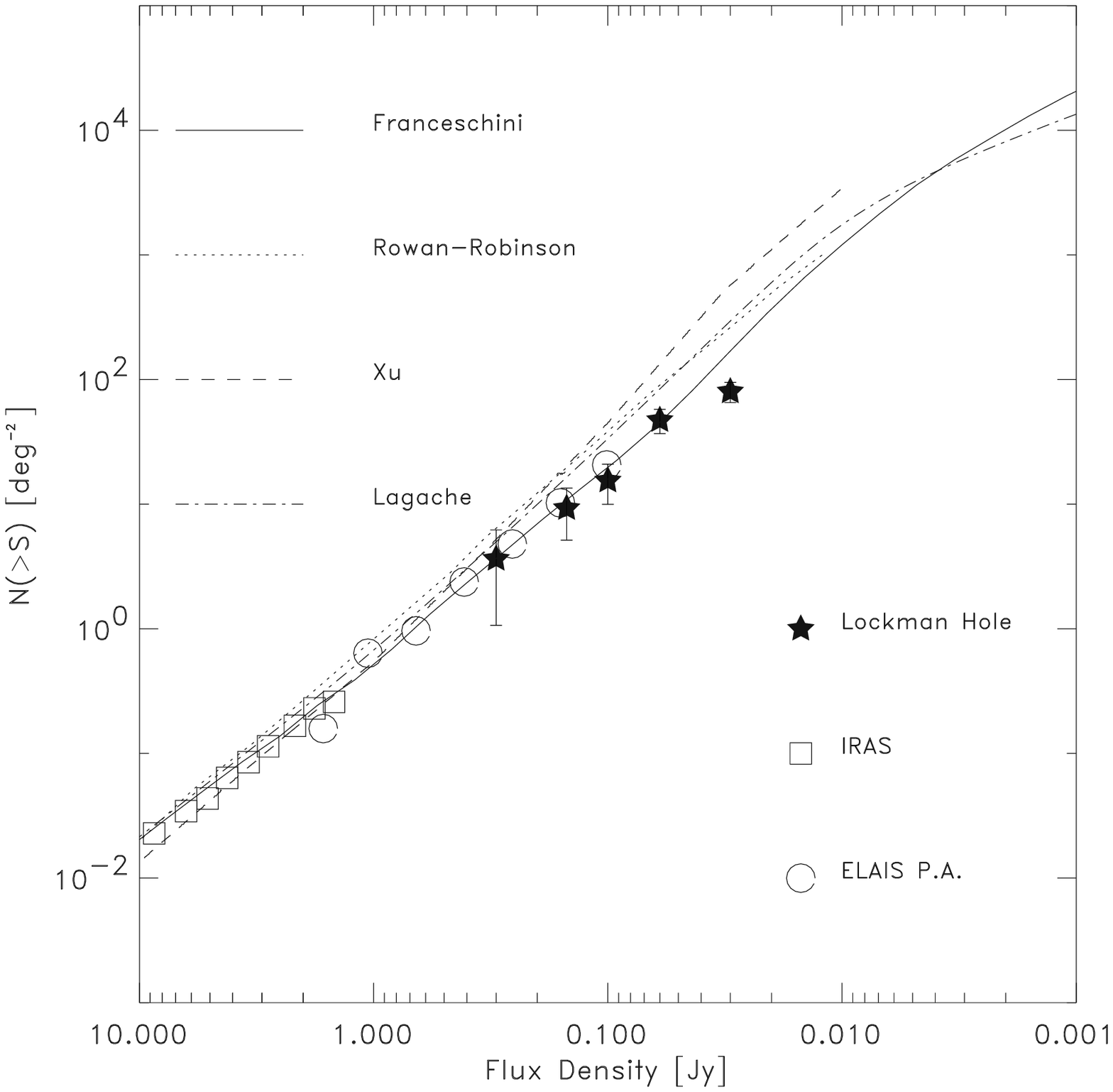, width=8cm} 
     \epsfig{file=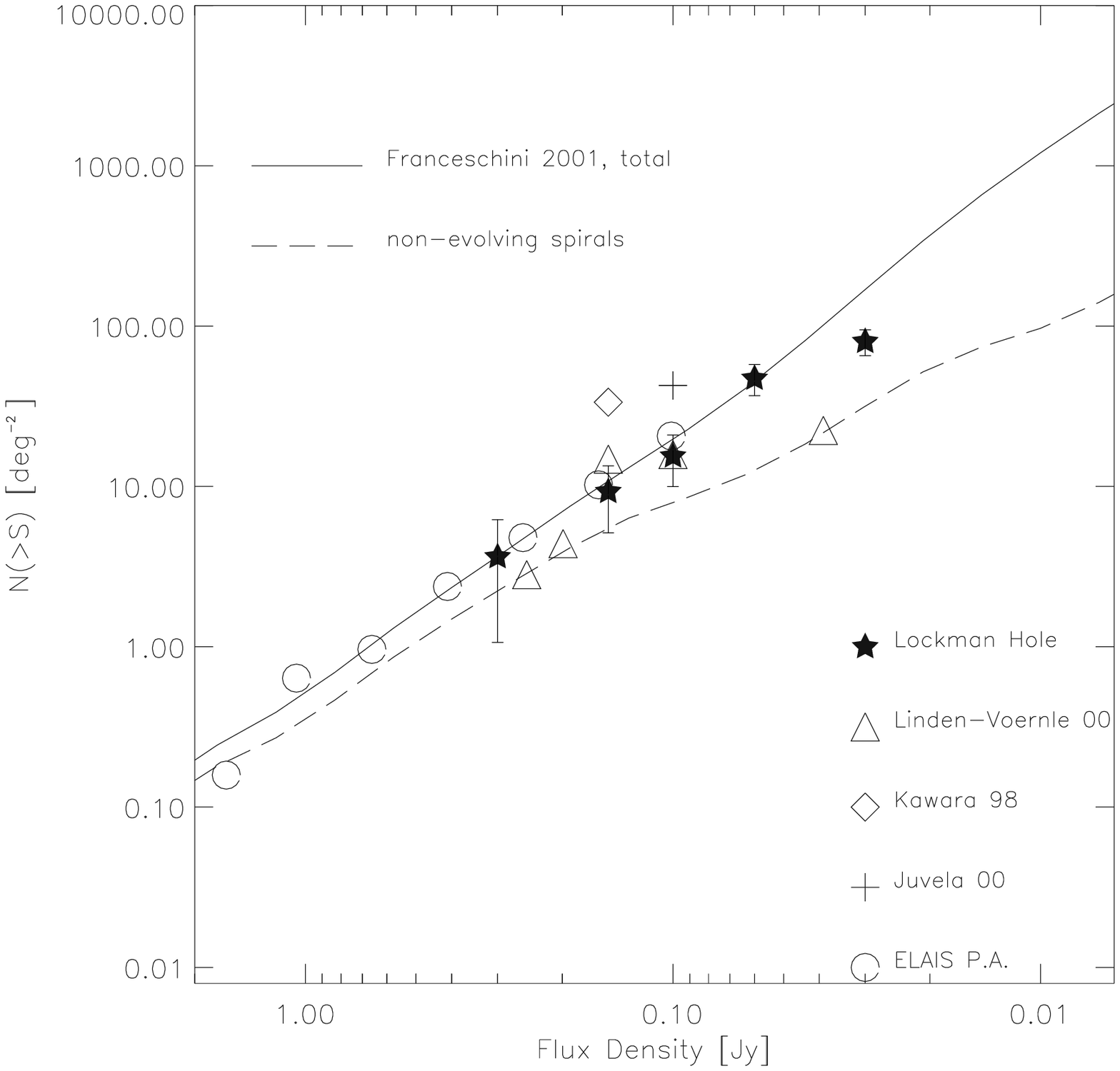, width=8cm} 
  \end{center} 
 \caption{Left panel (a): Integral counts at different flux levels. Our estimated values 
(starred symbols) are compared with data from other surveys: the preliminary analysis 
of ELAIS (open circles, Efstathiou et al., 2000), IRAS (open squares). A comparison is made
with predicted counts by Franceschini et al. (2001, solid line), Lagache et al. (2002,
dot-dashed line), Xu et al. (2001, dashed line) and Rowan-Robinson et al. (2001, dotted line).
 Right panel (b): comparison of our estimated integral counts with those published by 
Linden-Vornle et al. (2000, open triangles), Kawara et al. (1998, open diamonds) and 
Juvela et al. (2000, cross). The solid line is the model prediction by Franceschini et al.
(2001). We report as a dashed line the contribution of quiescient non-evolving spirals.} 
 \label{counts_i} 
 \end{figure*} 
 
\subsection{Source counts from the Lockman Hole LHEX 95 $\mu$m survey} 
\label{countslh}  

We concentrate in the present paper on discussing the statistical 
properties of the sample, like the source counts, confusion, and the
contribution of the detected sources to the cosmic far-IR background (CIRB).

\begin{figure*} 
  \begin{center} 
    \epsfig{file=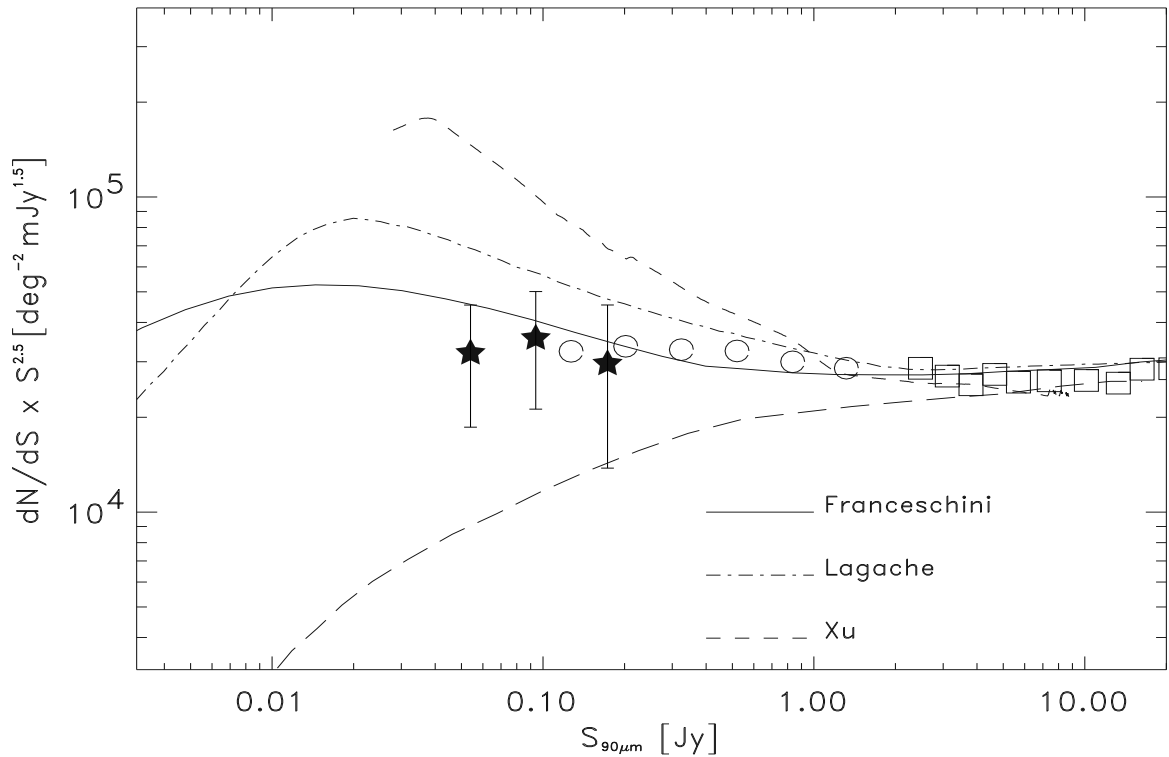,height=11cm,width=17cm} 
 \end{center} 
\caption{Differential 95 $\mu$m counts dN/dS normalized to the Euclidean law
($N \propto S^{-2.5}$). Symbols are the same as in Fig. \ref{counts_i}.
For comparison we report as a long-dashed line the contribution of non-evolving
spirals in the model of Franceschini et al. (2001). Our observed counts reveal
a significant excess above this curve at the faintest fluxes, supporting the
existence of an evolving population of IR galaxies.} 
\label{counts_d} 
\end{figure*} 
 
In the small area covered by the present study ($\sim$ 0.5 deg$^2$), 
we have computed the 95 $\mu$m source counts down to a flux level of 30 mJy
(in total 32 sources have been taken into account for this analysis). 
The integral counts have been obtained by weighting each single source 
for the effective area $A_{eff}$ corresponding to that source flux 
(as derived in Sect. \ref{compl}).  
The errors associated with the counts in each level have been computed as  
$\sqrt{\Sigma_i (1/A_{eff}^2(S_i))}$ (Gruppioni et al., 2002), 
where the sum is for all the sources with flux density 
$S_i$ and $A_{eff}(S_i)$ is the effective area. 
So the contributions of each source to both the counts and the associated
errors are weighted for the area within which the source is detectable.
These errors represent in any case the Poissonian term of the uncertainties, 
and have to be considered as lower limits to the total errors.

Our estimated values of the integral counts at different flux levels  
are plotted in Figure \ref{counts_i}a as starred symbols.  
Our results are compared here with those from other surveys: the preliminary 
analysis of the ISOPHOT ELAIS survey (Efsthatiou et al. 2000, open circles), 
and the counts derived from the IRAS 100 $\mu$m survey (open squares). 
Our data are in excellent agreement with these results in the 
flux range in common.  

The slope of the counts is $\alpha \sim 1.6$.

\begin{center}
\begin{table} 
\caption{Integral and differential source counts at 95 $\mu$m.} 
\begin{tabular}{|l|l| c c  c c c} 
\cline{1-2}
\cline{4-7} 
\multicolumn{2}{c|}{Integral}&&\multicolumn{3}{c}{Differential}\\
\cline{1-2}
\cline{4-7} 
S  & dN($>$S) 	&&	flux bin &bin center &number of sources& $dN/dS \times S^{2.5}$ \\
mJy   & deg$^{-2}$&&	mJy      & mJy       &detected in the bin & deg$^{-2}$mJy$^{1.5}$   \\
\cline{1-2}
\cline{4-7} 
30    & 80.1 & & 30-100   & 54  &23& $3.20e4\pm 1.3e4$\\
60    & 47.2 & & 60-150   & 94  &16& $3.56e4\pm 1.4e4$\\
100   & 15.5 & & 100-300  & 173 &6 & $2.96e4\pm 1.5e4$\\
\cline{4-7}
150   & 9.3 & & & & & \\
300   & 3.6  & & & & &\\
\cline{1-2}
\end{tabular} 
\label{ci} 
\end{table} 
\end{center}

A comparison of these integral counts with those published by Linden-Vornle 
et al. (2000),
Kawara et al. (1998), Juvela et al. (2000) is reported in Figure 
\ref{counts_i}b. We see
quite a substantial scatter in these data. We believe that our improved 
analysis
and careful check of all systematic and noise terms have produced a most
reliable outcome.
Obviously, our results are limited by the small survey area of about 0.5 
deg$^2$ and source statistics. However, we find encouraging our excellent
agreement with the source counts by Efstathiou et al. (2000), 
which are based on
a far larger survey area. We take this to indicate that our deeper survey
should not be biased too much by pathological clustering effects.

In Figure \ref{counts_d} we report the differential 95 $\mu$m counts dN/dS
normalized to the Euclidean law ($N\propto S^{-2.5}$), providing a statistically 
independent dataset to be compared with model predictions (the data are reported
in Table \ref{ci}).
A comparison is performed with modellistic 
differential counts by Franceschini et al. (2001), Xu et al. (2001) and 
Lagache et al. (2002).

The spatial distribution of the sources in our map of Fig. \ref{mosaic} is 
clearly non-random. We see in particular significant clustering in the East
sector of the map. The number of sources in the 4 quadrants are 7, 7, 13 
and 9 going in a clockwise order from the upper-right corner.
In this situation, the evaluation of the source confusion noise requires
some care.
Using the beam of the ISOPHOT C100 (see Sect. \ref{simul}) detector 
that has a FWHM of 45 arcsec, we have estimated that at the flux limit of 
20 mJy for each source in the map there are $\sim$90 independent cells.
This value is above the formal confusion limit, which is classically 
reached for a source areal density of $\sim$1/(30 independent beams) 
assuming Euclidean number counts (see Franceschini 1982 and
Franceschini 2000, section 8.3).
Around 20 mJy the Lockman 95$\mu$m counts are still close to the Euclidean
regime (see Fig. \ref{counts_d}).
In the map's quadrant with the highest number of sources, the areal density
is $\sim$1/(45 independent beams), still above confusion.
We conclude that our 95 $\mu$m map achieves a sensitivity close
to the confusion limit in its most crowded parts, but still should not be 
much affected by the confusion noise.
This is partly due to the significant incompleteness that our survey
suffers at the faintest flux limits. For an ideal complete survey, our
best-fit model implies a (3$\sigma$) confusion limit of $\sim$20 mJy
occurring at an areal density of 30 beams/source.
Kiss et al. (2001) have computed the $3\sigma$ confusion noise for ISOPHOT 
C100 90$\mu$m around $21\pm2$ mJy, while 
Matsuhara et al. (2000) report a value $\sim 30$ mJy . 
All these estimates confirm that our survey should not be confusion limited.

It may be instructive to compare these figures for the ISO 90 $\mu$m selection
with the confusion noise at longer wavelengths for observations with 
ISOPHOT C200 170$\mu$m.    From the analysis of the FIRBACK fields 
Kiss et al. (2001) report a $3\sigma$ value of $45\pm4$ mJy, while Dole et al.
(2001) estimate a $3\sigma$ confusion noise of 135 mJy.
This discrepancy is partly reconciled by considering that, in the computation 
of the confusion limit, Kiss et al. have used the default ISOPHOT PSF, while
Dole et al. have modeled the ISOPHOT 170$\mu$m beam, thus recovering the 
flux fraction stored in the external wings of an ideal source (which 
indeed is lost by using the standard PSF).
This confirms that the C100 imager is much less affected by confusion 
noise with respect to the C200's ($\sim20$ versus $\sim135$ mJy the respective
limits).
For this reason, in spite of the more severe problems related to the C100
data, the 90 $\mu$m ISOPHOT observations provide in principle a deeper view
and smaller error-boxes compared with longer wavelength observations.
 
\subsection{Source counts interpretation} \label{counts}  

We have compared in Figs. \ref{counts_i} and \ref{counts_d} our determined source 
counts with various modellistic estimates. The long dashed lines in Figs. \ref{counts_i}b 
and \ref{counts_d}, in particular, show a comparison with the 
predictions for a non-evolving source population: our observed counts
reveal a significant excess above these curves at the faintest fluxes, whose 
significance is better appreciated in Fig. \ref{counts_d} in terms of the independent
flux bins of the differential counts.
These results then confirm and substantiate earlier claims for the existence of an 
evolving population of IR galaxies, as previously identified in deep ISOCAM mid-IR
and ISOPHOT 175 $\mu$m counts. 

A comparison is made in Fig. \ref{counts_i}a and \ref{counts_d} with predicted counts 
by Lagache et al. (2002) (dot-dashed lines), Xu et al. (2001) (dashed line), 
and by Rowan-Robinson (2001) (dotted line in Fig. \ref{counts_i}a). 
These models  typically assume that the whole local galaxy population evolves 
back in cosmic time in source luminosity or number density
(with the exception of the Lagache model which accounts for both luminosity 
and density evolution).
All these curves fit well the bright IRAS counts, but tend to more or less exceed 
those at fainter fluxes.

\begin{figure} 
   \begin{center} 
     \epsfig{file=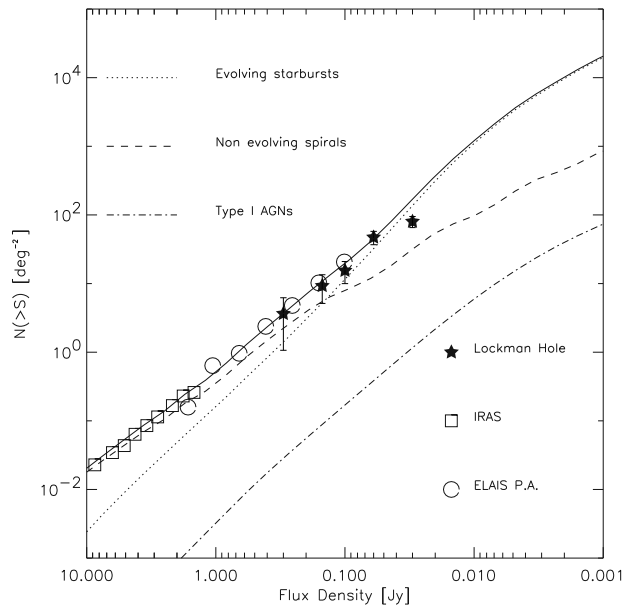, width=8.5cm} 
  \end{center} 
 \caption{95 $\mu$m integral counts compared with the different predicted IR populations
in the model of Franceschini et al. (2001). The main contributions come from non-evolving 
quiescent spirals (dashed line) and from a population of fast evolving sources
(dotted line) including starburst galaxies and type-II AGNs. 
A third component considered are type-I AGNs (dot-dashed line). The solid line
represents the total contribution.  } 
 \label{counts_mod} 
 \end{figure}

A better fit is provided by the multi-wavelength evolution model of Franceschini 
et al. (2001, hereafted F01) (solid lines). 
This model was designed to reproduce in particular the observed statistics (counts,
z-distributions, luminosity functions) of the ISOCAM mid-IR selected sources, but it
also accounts for data at other IR and sub-millimetric wavelengths. 
This model assumes the existence 
of three basic populations of cosmic sources characterized by different physical 
and evolutionary properties (their separate contributions are detailed in Figs.
\ref{counts_d} and \ref{counts_mod}). 
The main contributions come from non-evolving quiescent spirals (long dashed line in Figs. 
\ref{counts_d} and \ref{counts_mod}) and from a population of fast evolving sources
(dotted line in Fig. \ref{counts_mod}), including starburst galaxies and type-II AGNs 
(a third component 
considered - but always statistically negligible - are type-I AGNs, dot-dashed line 
in Fig. \ref{counts_mod}). 
The fraction of the evolving starburst population in the local universe 
is assumed to be $\sim$10 percent of the total, consistent 
with the local observed fraction of interacting galaxies. 

In this scenario, the active starbursts and the quiescent galaxies belong to the
same population. Each galaxy is expected to spend most of its lifetime in the 
quiescent state, but occasionally interactions and mergers with other galaxies 
put it in a short-lived (few to several $10^7$ years) active starbursting phase. 
The inferred cosmological evolution for the latter may be interpreted as an increased
chance to detect a galaxy during the active phase back in the past, possibly
due to a simple geometrical effect increasing the probability of interactions 
during the past denser epochs. 

F01 argued that this two-population model was needed to explain the particular
shape observed for the 15 $\mu$m counts. These keep a roughly euclidean slope
down to few mJy and quickly turn up at fainter fluxes (Elbaz et al. 1999; Gruppioni
et al. 2002; Fadda et al. 2003 in preparation). 
The two-population scheme (one evolving, one not) 
proposed by F01 allowed a best reproduction of these data.
Our new results about the galaxy counts at 95 $\mu$m seem to confirm this model 
against the predictions of single-population schemes.

If we now integrate our estimated 95 $\mu$m galaxy counts down to the limit
$S_{lim}=20$ mJy, we get a contribution $I$ of our resolved sources to the CIRB of
$I = \int_{\infty}^{S_{lim}} S\ dN/dS\ dS \simeq 2.0\ nW/m^2/sr$. The CIRB intensity
is rather uncertain at 100 $\mu$m due to the uncertain Zodiacal and
Galactic contributions, published values ranging from $\sim$11 (Lagache et al.
1999) to $\sim$22 $nW/m^2/sr$ (Hauser et al. 1998). Our estimated resolved contribution
then corresponds, respectively, 
to $\sim$18\% to $\sim$9\% of the CIRB intensity close to its peak wavelength.
By using the models of Rowan-Robinson et al. (2001) and Xu et al. (2001) 
we derive a contribution of $I\simeq 2.8\ nW/m^2/sr$ and $I\simeq 5\ nW/m^2/sr$,
respectively, to the CIRB at the limit of $S_{lim}=20$ mJy. These values 
are consistent with the spread observed in the counts predictions for the 
different models at 20 mJy (a factor 3 between the Franceschini and the Xu
models).

Deep confusion-limited maps with MIPS on SIRTF at 70 $\mu$m could not be compared
with the CIRB intensity (not directly measurable at this wavelength), while 
at 160 $\mu$m they are expected to be limited by confusion at $S_{160}\sim 80$ mJy
(Franceschini et al., 2001).
A significant improvement will require the substantially better spatial
resolution of the Herschel Space Observatory in 2007.

\section{CONCLUSIONS} 
We have developed a new procedure to clean and reduce deep survey data
obtained with the photo-polarimeter ISOPHOT C100 on board the Infrared
Space Observatory. These deep imaging data would have in principle
the advantage over longer wavelength ISOPHOT 175 $\mu$m observations 
of a better spatial resolution and a lower confusion noise, but there
use was limited by highly unstable detector background and responsivities.

Our procedure consists of a parametric algorithm fitting the signal time
history of each detector, and able to extract from it the background
level, to identify the singularities induced by cosmic-rays impacts
and by transient effects in the detectors, and 
to identify and extract real sky sources.

The source extraction process, completeness, the photometric and
astrometric accuracies of the final catalogues have been tested by us with 
extensive sets of simulations, by inserting into the real image sources
with known flux and position, and accounting for all the details of the 
procedure. The flux calibration was also verified by reducing with the
same technique C100 observations of calibrating stars.

We have tested our procedure by re-analysing data from a deep imaging 
survey performed at 95 $\mu$m with ISOPHOT C100 over a 
40$^\prime\times 40^\prime$ area within the Lockman Hole.
Within this area we detect thirty-six sources with S/N$>3$, making up a 
complete flux-limited sample for $S_{95 \mu m} \geq 100$ mJy. 
Reliable sources are detected, with decreasing but well-controlled 
completeness, down to $S_{95 \mu m} \simeq 20$ mJy.

These results provide us with the currently deepest far-IR image of the 
extragalactic sky. We estimate from it source counts down to a
flux of $\sim 30$ mJy, at which
limit we evaluate that of the order of 10\% to 20\% of the cosmic IR
background has been resolved into sources.

The 95 $\mu$m galaxy counts reveal a slope at $S_{95 \mu m} \le 100$mJy
quite steeper than that expected for a non-evolving source population.
These observed counts are consistent with those determined from ISO 
surveys at 
15 and 175 $\mu$m (Elbaz et al. 1999; Gruppioni et al. 2002; Puget et al. 
1999; Dole et al. 2001). The detailed shape of these counts constrains
the evolving population to dominate only below $\sim 100$mJy, whereas
at brighter fluxes the majority of the sources are expected to be 
massive spirals at moderate to low redshift. We will report on the
identifications and physical analyses of the 95 $\mu$m sources in 
separate papers (Aussel et al. 2003, in preparation; Fadda et al. 2003, in preparation; 
Rodighiero et al. 2003, in preparation).

\section*{Acknowledgments} 
 
We are grateful to M. Vaccari for his careful reading of a preliminary version 
of the paper.  
G.R. wants to thank F. Pozzi for her introduction to the problem of ISOPHOT calibration.  
We thank the anonymous referee for his helpful suggestions that improved
the paper.
 
This work was partly supported by the "POE" EC TMR Network Programme 
(HPRN-CT-2000-00138).

\appendix 

\section{THE LINEARITY CORRECTION} \label{linearity} 

One of the main aspects of the PHOT-C detector is its non-linear 
response to the incoming flux, which requires a suited correction that has to be  
carefully derived. We have tried to follow a different approach 
from that of the standard PIA pipeline. 
An accurate analysis of the good stability of the detectors 
with time, enable us to construct a linearity correction starting from the dataset 
available for the S1 ELAIS field (for a total area of $\sim$ 4.5 deg$^2$) 
and compare the result to the PIA internal calibration.  
 
In order to obtain such correction, we have divided the data in 121 
Volts intervals (from -1.2 to 1.2, covering the full dynamical range 
of the detector). Inside each step of voltage we have computed the median of the 
differences in ADU/gain/s and normalized to the median over the whole 
dynamical range. This means that the zero point of our correction is 
referred to the voltage range where the vast majority of the observed 
data in the ELAIS S1 field is located ($-0.8<V<-0.4$). 
We have taken into account for the computation only data around the 
background (to neglect the influence of deep and high transients in the 
final correction), discarding every readout corresponding to distructive 
readings, to the first two points of each ramp, to the sources and to the 
saturated readings. 
  
\begin{figure} 
  \begin{center} 
  \epsfig{file=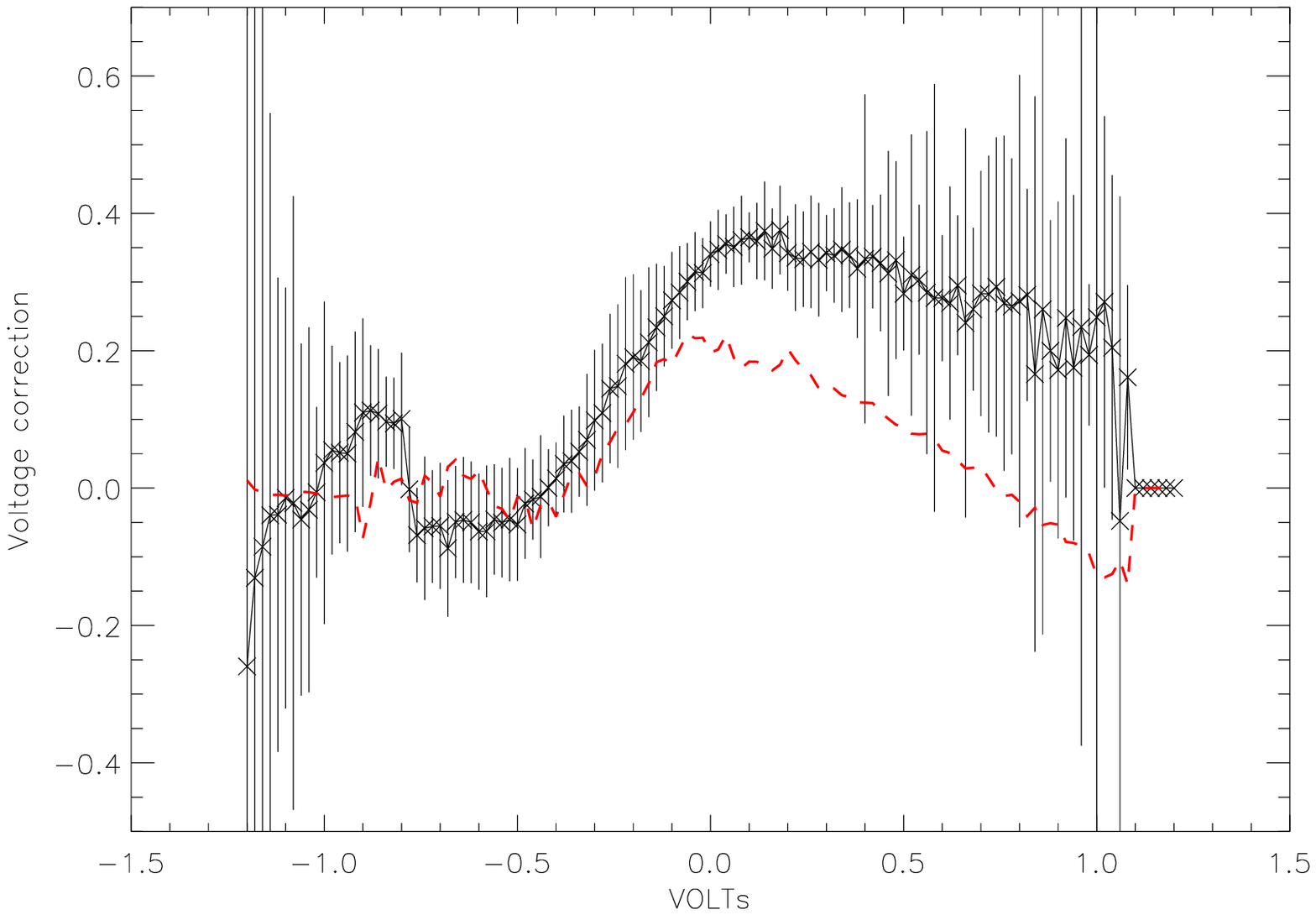, width=9cm} 
  \end{center} 
\caption{Comparison between PIA linearity correction (dashed line) and the 
correction we derived from data in the ELAIS S1 field, as a function of 
the voltage level for one ISOPHOT detector pixel.  
The rms we get for each volt level are reported.  
The corrections are percentual.} 
\label{lin} 
\end{figure}

The result for an example pixel is shown in Figure \ref{lin},  
where PIA standard correction (dashed line) is compared with ours, 
as a function of voltage. The corrections reported are percentual. 
There is a quite good agreement at $V<0$ (the deviation is of the order 
of 10$\%$), considering the larger quantity of data used in the computation 
of the PIA correction. 
 
\begin{figure} 
  \begin{center} 
     \epsfig{file=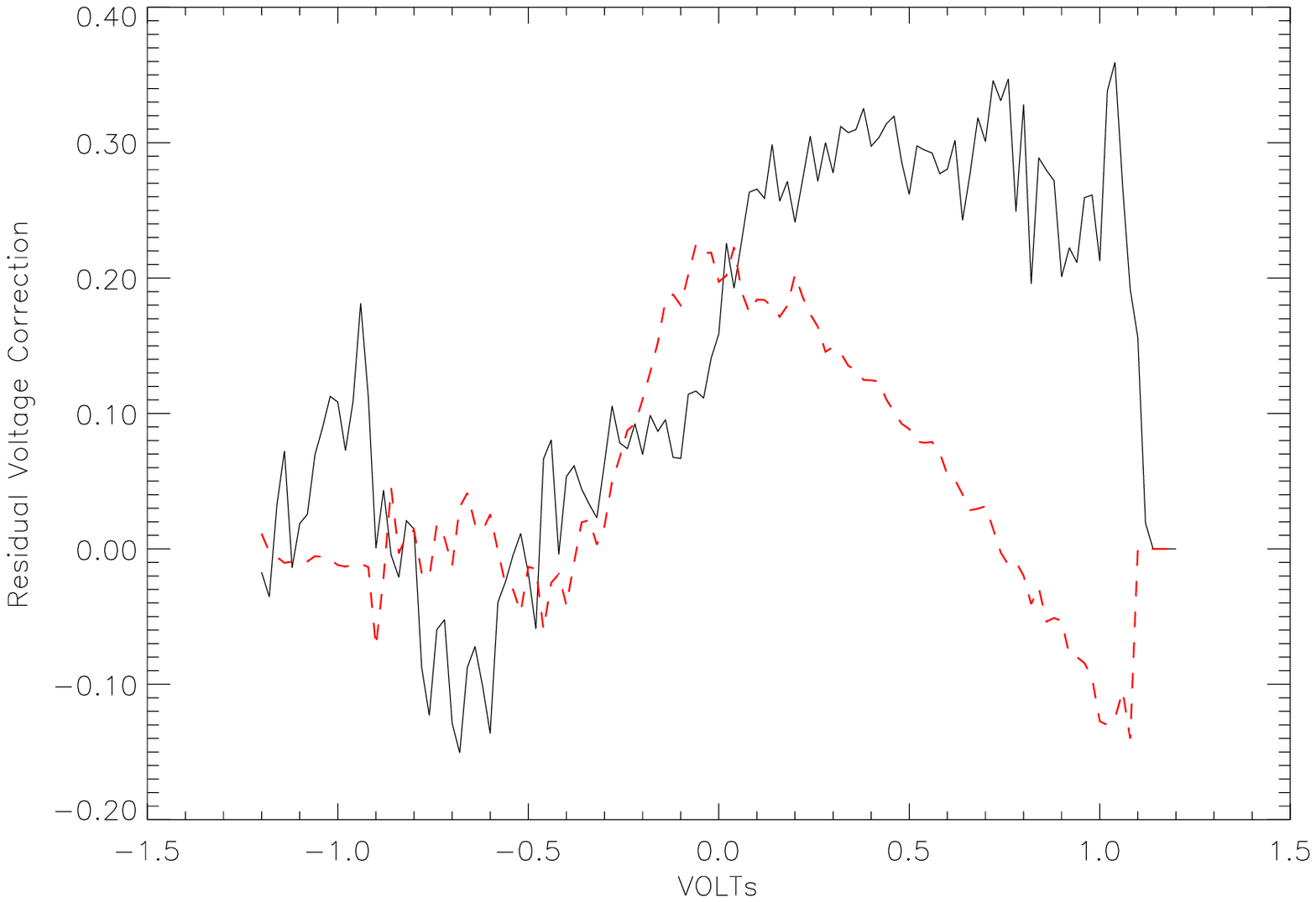, width=9cm} 
  \end{center} 
\caption{The PIA standard correction (dashed line, as in Fig. 4) is compared 
with our $residual$ correction (solid line). See text.} 
\label{lin2} 
\end{figure}

To check the consistency and the quantitative effects of the two 
corrections, we have estimated the residual correction we get after 
the data have been processed with PIA. In other words: we have derived 
with the same prescriptions our correction from data already PIA 
corrected. Such "residual" correction is shown if Figure \ref{lin2} 
(continuous line) and compared with the usual PIA percentual correction 
(dashed line, as in Figure \ref{lin}).  
It is clear that at $V>0$ the PIA calibration is underestimated, and 
that we need to apply a further second order correction for a better 
linearity at the higher voltages (that we took into account).  
 
We have considered the possibility of a further dependence of the 
linearity on the ramp position. 
In a similar way to that of the voltage dependence, we have thus construct  
a second order "geometrical" correction from the data, normalized to the 
central region of the ramp, that allows to correct any residual  
non-linearity (a mean correction for the dependence of the ramp 
index erases completely the effects on the background, but can 
leave residuals on the stronger sources).

\section{FLUX CALIBRATION} 
\label{calib}

In order to get a good flux calibration, we have tried to compute a new 
estimate of the detector pixels responsivities.  
We mean with responsivities those 
factors that convert digital units [ADU/gain/s] to physical flux units [mJy]. 
We made use of a wide set 
of internal lamps, the so called FCS (Fine Calibration Sources,  
ISOPHOT Handbook, Laureijs et al., 2002). 
In particular we have taken into account all FCS measurements 
associated to raster observations in the ELAIS fields, in the Lockman Hole 
and to other minor mini-rasters of point sources,  
in order to cover an extended range  
of voltage levels. For every raster here considered, observed in the P22 mode, 
there are two FCS measurements available: the first ($FCS_1$)  
taken before the science  
observation, and a second ($FCS_2$) at the end of the observation.  
As discussed in section \ref{red}, we are confident about the stability of the 
detectors as a function of time. This justifies our attempt to compute  
the responsivities from a set of observed data. 
 
We have reduced the FCS data following the same procedures described 
in this paper for raster science observations.  
However, we have previously applied the dark-current correction  
and the reset-interval correction, both part of the 
standard PIA package (Gabriel \& Acosta-Pulido, 1999; Schulz et al., 2002). 
Our good statistics of internal calibrators enable us to reject all bad  
observations (mostly due to the massive presence of cosmic rays),  
and take into account only those for which the fitting algoritm has not failed. 
The main problem when dealing with FCS data is their short total exposure time  
(usually 1024 readouts), that prevents to determine a correct value of the 
stabilization level and may translate in a understimate of the flux. 
 
\begin{figure} 
  \begin{center} 
    \epsfig{file=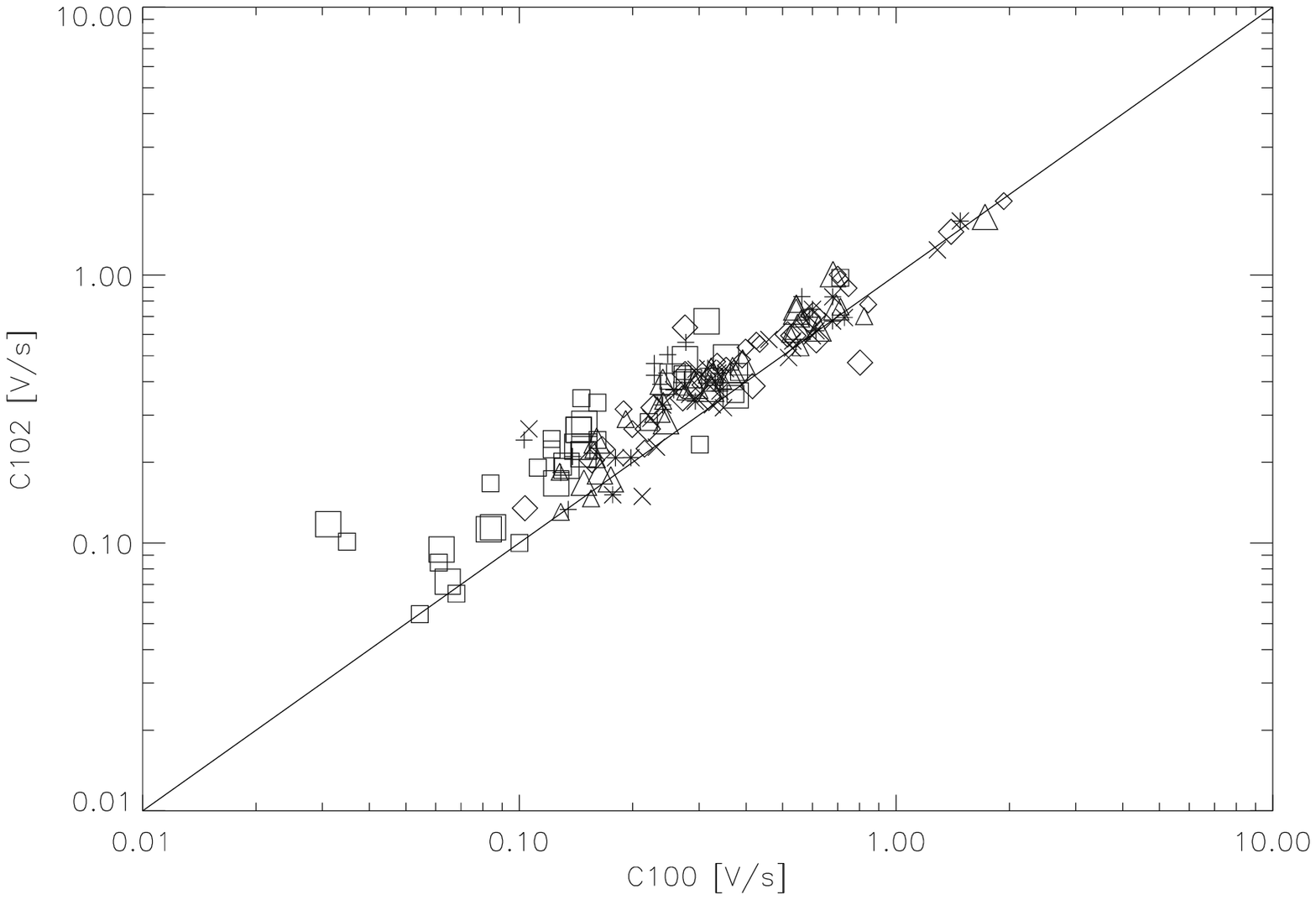, width=9cm} 
\caption{The figure shows the correlation between   
$FCS_1$ and $FCS_2$ reduced with our pipeline.  
They are well correlated over the whole  
range of voltages considered, and for all pixels. 
The values reported are in units of Volts/s and represent  
the median values of every FCS observation. 
Different symbols refer to different pixels. The solid line represent  
the 1 to 1 relation.} 
\label{lamps} 
 \end{center} 
\end{figure}

Once the lamps have been reduced, we proceed to the statistical  
analysis for the calculation of responsivities. Figure \ref{lamps} shows  
that with our reduction  
$FCS_1$ and $FCS_2$ are well correlated over the whole range of  
voltages considered, and 
for all pixels. The values reported are in units of Volts/s and  
represent the median 
values of every pointing (each FCS is composed of a single pointing). 
Different symbols refer to different pixels.  
The solid line represent the 1 to 1 relation. 
There seems to be a slight tendency for $FCS_2$ values to be 
statistically higher than $FCS_1$ values, in particular at lower fluxes. 
This could be due to the long integration times that divides the two 
lamps in each observation.
After the science observation the detector is still 
quite hot, and a residual bias level could rest in the slowly varing memory  
of each pixel, offseting upward the stabilization level of each FCS2.  
This effect might be erased, or at least reduced, with the avalaibility  
of longer FCS exposure times, that would improve the determination of  
the stabilization. 
 
In order to compute the responsivities, we chose to use the average value 
between each pair of $FCS_1$ and $FCS_2$. In the following we will call these  
values $V_{FCS}$ [Volts/s]. $V_{FCS}$ are related to the responsivities via the 
equation  
 
\begin{equation} 
RESP(i)=V_{FCS}(i)*K(i)/P_{FCS}(i) 
\label{resp} 
\end{equation} 
 
where $RESP$ are the responsivities,  
$K$ is a costant which depends only on the pixel and accounts for the 
illumination matrix and other instrumental effects of the detector, 
$P_{FCS}$ are the Inband fluxes (which represent the expected fraction  
of energy filtered through the system detector+filter, and collected as output signal on the detector), expressed in units of Watts. The $i$ index  
indicates the dependence of each variable on the different nine pixels. 
The previous equation is a syntax simplified version of the formula reported by 
Schulz et al. (2002) and in the ISOPHOT Handbook. 
 
In order to compute the inband fluxes we have used the standard PIA power 
curve (Schulz et al. 2002), derived from a careful and complete analysis of 
external calibrators (stars, planets and asteroids). This calibration curve allows to correct the given heating powers (the total energy emitted from the FCS), to inband fluxes. 
 
We can rewrite equation \ref{resp} as: 
 
\begin{equation} 
V_{FCS}(i)=RESP(i)*P_{FCS}(i)/K(i) 
\label{v_fcs} 
\end{equation} 
 
and compute for every pixel the best linear fit between the two independent  
variables $V_{FCS}(i)$ and $P_{FCS}(i)/K(i)$ (each observation of our selected 
calibration data-set gives a point in this linear relation). The slope  
of this relation ($RESP(i)$) represents the responsivities.  
It is clear that the responsivities we get with this procedure are only a  
function of the pixel, and are assumed to be constant through the temporal 
history of the satellite. The values we derived are consistent with that  
of the standard PIA values.
 
The responsivities function of the pixels represent our best flat-field.

Finally, we can use our responsivities to calibrate our maps (to convert 
the signal from ADU/gain/s to flux units of mJy/pixel). 
 
In order to check our calibration and look for any further  
physical scale factor (the absolute calibration) we have studied and reduced  
a few external calibrators (see discussion in Section \ref{flux1}).  
The good agreement  
we found at different flux levels and the intrinsic uncertainties on previous 
95 $\mu$m measurements, made us confident to use the responsivities 
without appling any offset correction. 
We have estimated the photometric errors of our calibration throughout  
simulations and found they are of the order of 20-30 percent.

\section{DETAILS ON THE SIMULATION PROCESS} \label{simul_app}

To correctly simulate a source on a raster map and to determine its  profile along  
the pixel time history, we make use of the ISOPHOT C100 PSF and of the projection 
algorithm. 
PHOT readout frequency is higher with respect to ISOCAM (1/32 seconds versus 2 
seconds), the pixel size is greater (45 x 45 arcsec) and the distance between 
adjacent raster pointings is quite large (of the order of half detector array, 
69 arcsec, in the case of the Lockman Hole, greater for other surveys like ELAIS). 
The combination of these effects requires the simulation of an  high resolution  
source profile. 
In the  ISOCAM procedure, the predicted flux level of a source was projected on every 
raster pointing and  assumed to be there constant.  For PHOT we have created 
a second raster structure with astrometric informations on each readout along the 
time history,  including those where the satellite moves from a raster 
position to the other (not on target readouts). 
The projection of the PSF on these more gridded series of pointings and the  
lecture  of  these levels projected on every readout,  is able to reproduce  the 
exact profile of how the detector should see a source (if its response would be linear 
to the incoming flux).  
The detail description of this profile is mainly important between two raster pointings, 
when the detector moves on the sky. Given the size of the PHOT pixels, the detector 
starts to see a source  when it is still moving and not yet positioned in a raster pointing. 
If we do not take into account this flux recorded by the detector (when the source  
enters and leaves a pixel) we could underestimate the total flux of the source. 
This correction is crucial when simulations are used to compute the corrections 
to the fluxes of the detected sources.   
In the following step, the projection algorithm we use takes into account the 
transient behaviour  of the detector, and "models" the simulated source profile  
along the time history in order to introduce the instrumental effects and obtain 
sources very similar to the real ones.  
 
Our "simulator" is very efficient, as it can describe sources independently of their 
position inside the pixel.

\label{lastpage}

\end{document}